\documentclass[prl,aps,showpacs,superscriptaddress,twocolumn]{revtex4}
\pdfoutput=1
\usepackage{color} 
\usepackage{times}
\usepackage{amssymb} 
\usepackage{amsmath} 
\usepackage{graphicx}
\usepackage[english]{babel}
\usepackage{epstopdf}
\usepackage{mathrsfs}
\usepackage{textcomp}
\usepackage{makeidx}
\usepackage{amsthm}
\usepackage{amsfonts}
\usepackage{amssymb} 
\usepackage{hyperref}
\usepackage{soul}

\newcommand{\ket}[1]{\vert#1\rangle}

\newcommand{\ketbra}[2]{\vert#1\rangle\langle#2\vert}

\newcommand{\opt}[4]{\hat{#1}_{#3}^{#4}(#2)}
\newcommand{\op}[3]{\hat{#1}_{#2}^{#3}}
\newcommand{\opd}[3]{\check{#1}_{#2}^{#3}}

\begin{document}

\title{Non-Markovianity criteria for open system dynamics}

\author{N. Lo Gullo}
\affiliation{Department of Physics, University College Cork, Cork, Republic of Ireland}
\affiliation{Quantum Systems Unit, Okinawa Institute of Science and Technology and Graduate University, Okinawa, Japan}
\affiliation{Dipartimento di Fisica e Astronomia, Universit\`a degli Studi di Padova, Padova, Italy}
\author{I. Sinayskiy}
\affiliation{Quantum Research Group, School of Chemistry and Physics, University of KwaZulu-Natal, Durban, 4001, South Africa}
\affiliation{National Institute for Theoretical Physics (NITheP), KwaZulu-Natal, South Africa}
\author{Th. Busch}
\affiliation{Department of Physics, University College Cork, Cork, Republic of Ireland}
\affiliation{Quantum Systems Unit, Okinawa Institute of Science and Technology and Graduate University, Okinawa, Japan}
\author{F. Petruccione}
\affiliation{Quantum Research Group, School of Chemistry and Physics, University of KwaZulu-Natal, Durban, 4001, South Africa}
\affiliation{National Institute for Theoretical Physics (NITheP), KwaZulu-Natal, South Africa}

\date{\today}
\begin{abstract}
A universal definition of non-Markovianity for open systems dynamics is proposed. 
It is extended from the classical definition to the quantum realm by showing 
that a `transition' from the Markov to the non-Markov regime
occurs when the correlations between the system and  the environment, 
generated by their joint evolution, can no longer be neglected. The suggested definition is based on the comparison between measured correlation
functions and those built by assuming that the system is in a Markov regime
thus giving a figure of merit of the error coming from this assumption. It is shown that the knowledge of the dynamical map and initial condition of the system is not enough to fully characterise the non-Markovian dynamics of the reduced system. The example of three exactly solvable models, i.e. decoherence and spontaneous emission of the qubit in a bosonic bath and decoherence of the photon's polarization induced by interaction with its spacial degrees of freedom, reveals that previously proposed Markovianity criteria and measures which are based on dynamical map analysis fail to recognise non-Markov behaviour. 
\end{abstract}

\maketitle

{\it Introduction } -
The quest for a mathematical description of open quantum system dynamics
has lately intensified~\cite{open}. 
The reason has to be sought in the application of this theory
to problems such as thermalization of quantum systems~\cite{therm},
transport in non-equilibrium settings~\cite{noneq}, 
dynamics of quantum systems in noisy environments 
with non trivial correlations~\cite{noise}, and
the emergence of irreversibility from microscopic description 
(as first studied by Lindblad~\cite{lind}), to mention a few.
From a theoretical perspective the study of a quantum system requires the knowledge of
the mean values of its observables {\it and} their correlation functions.
The latter are often linked to measurable (and important) quantities such as the decay
rate of a system, generalized susceptibilities and scattering properties of a field.
A powerful tool for the calculation of the $n$-point correlation functions is the so-called Quantum Regression Theorem (QRT)~\cite{lax};
it states that the evolution equations of the correlation functions can be related to 
the evolution equation for the mean values of the observables themselves
in the spirit of the Onsager hypothesis for classical systems~\cite{onsager}.
However, the QRT has been proven to not be the quantum 
generalization of the Onsager hypothesis~\cite{noqrt}, which means that,
in general, there is no other way of calculating the correlation functions
but starting from a microscopic model and going through the Heisenberg evolution
of operators; this task is often non trivial.

A class of open quantum systems exists, whose mathematical description
is fairly simple and for which it is possible to derive
a master equation for the time evolution of the density matrix of 
Gorini-Kossakowski-Sudarshan-Lindblad (GKSL) form~\cite{GKS,L}.
This equation is said to generate a Markovian dynamics because for
diagonal density matrices it is equivalent to the Pauli equation~\cite{pauli}
and the solution to the Pauli equation is known to fulfill the Markov assumption
for classical systems.
One of the properties of the GKSL master equation is that it
generates equations of motion for the correlation functions
which have the same structure of the equation of motion for the mean values,
{\it i.e.}, the QRT holds true for systems described by this equation~\cite{gard}.

The general evolution of an open system is described
by an equation which can be different from the GKSL form and in
this case it is customary to talk about non-Markovian master equations,
implicitly assuming that the relation ``GKSL form $\leftrightarrow$ Markovian dynamics''
holds.
However, unlike in the classical case,
in the quantum case there is no clear definition of Markovian systems, and it has recently been argued~\cite{class} that for a quantum system it is not 
possible to define Markovian processes in analogy with the classical case.
This is due to the difficulty in defining conditional quantities in quantum mechanics
because of the perturbation induced by measurements.
In order to better classify the different types of dynamics
and distinguish between Markovian and non-Markovian evolution
different measures of non-Markovianity have been proposed~\cite{breuer, div1, div2,volmap,comp1,comp2}. 
These approaches 
are based on distinguishability and divisibility of the corresponding maps. However, in the corresponding classical case these properties of the maps are not sufficient to guarantee the Markovianity of the process ~\cite{hanggi1,hanggi2}.

The aim of this letter is to give a clear and experimentally testable definition of the Markov regime
for the dynamics of an open system. It is important to mention that the proposed definition is valid for quantum and classical dynamics. The definition of the Markov process will be based on
an appropriate 
reformulation of the Markov assumption for classical stochastic processes~\cite{class}.  As a consequence of our definition we show why any definition of the Markov process based on the analysis of the properties of the dynamical maps only leads to inconclusive results. 
We will also show that our definition of the Markovian regime
has implications for the study of fluctuation relations in
open systems~\cite{nonGKSL,FRLindblad}.

{\it Markov assumption for classical stochastic systems }- 
The description of a classical stochastic system
is given, in general, by a set of probability distribution functions
$ p^{(n)}(x_n,t_n;\;\cdots;\;x_1,t_1)$, for $(t_n\ge\cdots\ge t_1)$. These allow to calculate 
mean values and multi-times correlation functions as
$\langle f_n(X)\cdots f_1(X)\rangle(t_n,\cdots,t_1)=\int dx_n\cdots dx_1\; p^{(n)}(x_n,t_n;\;\cdots;\;x_1,t_1)\; f_n(x_n)\cdots f_1(x_1) $,
where $x_n$ is the value of the stochastic variable $X$ of the system
at time $t_n$
and the $f$'s are functions of the stochastic variable $X$ 
relative to some system's observable
(e.g., energy, temperature, components of the velocity).
There is a particular class of systems whose mathematical description
is particularly simple; they are the ones for which the so-called Markov
assumption holds true \cite{Feller}. 

The Markov assumption for classical systems is stated through
the conditional probabilities and reads

\begin{equation}
p_{n | k}({\bf x}_{n+k,k+1}\; |\;{\bf x}_{k,1})=p_{n | 1}({\bf x}_{n+k,k+1} | x_k,t_k),
\end{equation}

\noindent where we have defined ${\bf x}_{j,i}=\{x_j,t_j;\;\cdots;\;x_i,t_i\}$ and 
$p_{n | k}({\bf x}_{n+k,k+1}\; |\;{\bf x}_{k,1})=p_{n+k}({\bf x}_{n+k,1})/p_{k}({\bf x}_{k,1})$.
\noindent
This in turn implies that once the two-points conditional probability density distribution and 
the initial conditions are given one can evaluate {\it any} $n$-point (in time) correlation
functions. Hence, the system is {\it fully characterized}.

Let $p^{(1)}_0(x_0,t_0)$ be the initial probability density, 
then we can formally write~\cite{hanggi1}

\begin{equation}
p^{(n)}({\bf x}_{n,1})= \int d \tilde x\;G^{(n)}({\bf x}_{n,1}| \tilde x,\tilde t)\,p^{(1)}_{0}(\tilde x,\tilde t),
\end{equation}

\noindent where the $G^{(n)}({\bf x}_{n,1}| \tilde x,\tilde t)$ are the Green's functions
that solve the initial problem.
H\"anggi and Thomas showed~\cite{hanggi1,hanggi2} that it is always possible 
(Markov and non-Markov case)
to construct a class of propagators $G^{(n)}({\bf x}_{n,1}| \tilde x,\tilde t)\,p(\tilde x,\tilde t)$ 
(for each $n$) with the semi-group property and which are independent 
from the initial conditions.

Nevertheless these Green's functions cannot be interpreted as
conditional probability distribution densities.
As pointed out in~\cite{hanggi1}, 
the one point propagators $G(t|t_1)$ of each semi-group are the 
conditional probabilities of a Markov process which has the same single-event probabilities 
$p(t)$, {\it but different multivariate probabilities $p^{(n)}$},
as the non-Markov process under consideration.
Therefore, the knowledge of the single particle propagator tells nothing about the 
higher order correlation functions. The same holds true for higher order propagators and
this is the crucial starting point for our discussion.
Markovianity in a quantum framework is often related to the divisibility, or to the semi-group
property, of the so-called dynamical map, and measures of Markovianity have been suggested based on this property \cite{div1,div2}.
However, in the classical case there is no such direct implication.

{\it Multivariate correlation functions and definition of the Markov regime for quantum dynamics} -
Let us first consider a quantum system interacting with an environment.
The joint dynamics is given by a unitary operator $\op{U}{t_0}{t}=e^{-\imath\op{H}{}{} (t-t_0)}$,
where $\op{H}{}{}$  is the total Hamiltonian $(\hbar=1)$ , and
the dynamical map is defined as $\Phi_{t_0}^{t}: \opt{\rho}{t_0}{S}{}=\text{Tr}_E\big[\opt{\rho}{t_0}{}{}\big] \rightarrow \opt{\rho}{t}{S}{}=\text{Tr}_{E}\big[\op{U}{t_0}{t}\opt{\rho}{t_0}{}{}(\op{U}{t_0}{t})^{\dag}\big]$.
Here $\opt{\rho}{t_0}{}{}$ is the initial density operator of the total system
and the indices $S$ and $E$ refer to system and 
environment, respectively.
Finally, let us specify a reference state for the environment, which we will 
choose to be its initial state 
$\op{\rho}{E}{}=\text{Tr}_S[\opt{\rho}{t_0}{}{}]$.

To define the Markov regime, we
recast the multivariate correlation functions into a form more useful for our purpose.

For this we consider the two-point correlation function
$\langle\opt{o}{t+\tau}{1}{}\opt{o}{t}{2}{}\rangle$, where the operators $\op{o}{i}{}$ act only 
on the system's Hilbert space.
By using the cyclic property of the trace together with the expression 
$\opt{o}{t}{i}{}=e^{\imath\op{H}{}{}t}\op{o}{i}{} e^{-\imath\op{H}{}{}t}$ and defining
$S_{t_1}^{t_2}[\;\circ\;]=\op{U}{t_1}{t_2}\circ(\op{U}{t_1}{t_2})^{\dag}$, the correlation function can be recast in the form 
\begin{equation}
 \label{eq:TPCF}
 \langle\opt{o}{t_2}{2}{}\opt{o}{t_1}{1}{}\rangle=
 \text{Tr}_S\left[\op{o}{2}{}\text{Tr}_E\left[S_{t_2}^{t_1}\left[\,\op{o}{1}{}S_{t_0}^{t_2}[\opt{\rho}{t_0}{}{}]\,\right]\;\right]\right].
\end{equation}

The functionals
$\mathcal{P}$ and $\mathcal{Q}$ are defined such that 
$\mathcal{P}[\op{O}{}{}]=\text{Tr}_E[\op{O}{}{}]\otimes\op{\rho}{E}{}$ and $\mathcal{Q}=\mathcal{I}-\mathcal{P}$, where $\op{O}{}{}$ is an operator acting on the total Hilbert space
(system plus environment) and $\mathcal{I}[\op{O}{}{}]=\op{O}{}{}$ is the identity 
functional.
Inserting three identities $\mathcal{I}=\mathcal{P}+\mathcal{Q}$ into Eq.~\eqref{eq:TPCF}, we obtain the expression

\begin{widetext}
\begin{eqnarray}
\label{eq:multi}
\langle\opt{o}{t_2}{2}{}\opt{o}{t_1}{1}{}\rangle=\;&\text{Tr}_S\left[\op{o}{2}{}\,\text{Tr}_E\left[\,\mathcal{P}\,S_{t_1}^{t_2}\,\op{o}{1}{}\,\mathcal{P}\,S_{t_0}^{t_1}\,\mathcal{P}\,\opt{\rho}{t_0}{}{}\;\right]\right]+\text{Tr}_S\left[\op{o}{2}{}\,\text{Tr}_E\left[\,\mathcal{P}\,S_{t_1}^{t_2}\,\op{o}{1}{}\,\mathcal{P}\,S_{t_0}^{t_1}\,\mathcal{Q}\,\opt{\rho}{t_0}{}{}\;\right]\right]\\
+&\text{Tr}_S\left[\op{o}{2}{}\,\text{Tr}_E\left[\,\mathcal{P}\,S_{t_1}^{t_2}\,\op{o}{1}{}\,\mathcal{Q}\,S_{t_0}^{t_1}\,\mathcal{P}\,\opt{\rho}{t_0}{}{}\;\right]\right]+\text{Tr}_S\left[\op{o}{2}{}\,\text{Tr}_E\left[\,\mathcal{P}\,S_{t_1}^{t_2}\,\op{o}{1}{}\,\mathcal{Q}\,S_{t_0}^{t_1}\,\mathcal{Q}\,\opt{\rho}{t_0}{}{}\;\right]\right]\nonumber\\
+&\text{Tr}_S\left[\op{o}{2}{}\,\text{Tr}_E\left[\,\mathcal{Q}\,S_{t_1}^{t_2}\,\op{o}{1}{}\,\mathcal{P}\,S_{t_0}^{t_1}\,\mathcal{P}\,\opt{\rho}{t_0}{}{}\;\right]\right]+\text{Tr}_S\left[\op{o}{2}{}\,\text{Tr}_E\left[\,\mathcal{Q}\,S_{t_1}^{t_2}\,\op{o}{1}{}\,\mathcal{P}\,S_{t_0}^{t_1}\,\mathcal{Q}\,\opt{\rho}{t_0}{}{}\;\right]\right]\nonumber\\
+&\text{Tr}_S\left[\op{o}{2}{}\,\text{Tr}_E\left[\,\mathcal{Q}\,S_{t_1}^{t_2}\,\op{o}{1}{}\,\mathcal{Q}\,S_{t_0}^{t_1}\,\mathcal{P}\,\opt{\rho}{t_0}{}{}\;\right]\right]+\text{Tr}_S\left[\op{o}{2}{}\,\text{Tr}_E\left[\,\mathcal{Q}\,S_{t_1}^{t_2}\,\op{o}{1}{}\,\mathcal{Q}\,S_{t_0}^{t_1}\,\mathcal{Q}\,\opt{\rho}{t_0}{}{}\;\right]\right]\nonumber,
\end{eqnarray}
\end{widetext}

\noindent (in order to simplify the notation we dropped square brackets and assume that 
functionals act on {\it everything} on their right hand side). By means of the definitions of the functionals $\mathcal{P}$ and $\Phi$~\cite{calc} 
we can rewrite the first term in Eq.~(\ref{eq:multi}) as
\begin{equation}
\label{eq:mark}
\begin{aligned}
\text{Tr}_S&\left[\op{o}{2}{}\,\text{Tr}_E\left[\,\mathcal{P}\,S_{t_1}^{t_2}\,\op{o}{1}{}\,\mathcal{P}\,S_{t_0}^{t_1}\,\mathcal{P}\,\rho(t_0)\;\right]\right]\\
&\qquad=\text{Tr}_S\left[\op{o}{2}{}\,\Phi_{t_1}^{t_2}\,\op{o}{1}{}\,\Phi_{t_0}^{t_1}\,\rho_S(t_0)\right],
\end{aligned}
\end{equation}
and it is easy to extend this expression to a generic $n$-points correlation function to obtain 
formulae analogous to that of Eq.~\eqref{eq:mark}.
It is well known that if the infinitesimal generator of the system's dynamics is 
of the GKSL form~\cite{open, gard}, then the relation 
$\langle\opt{o}{t_2}{2}{}\opt{o}{t_1}{1}{}\rangle=\text{Tr}_S\left[\op{o}{2}{}\,\Phi_{t_1}^{t_2}\,\op{o}{1}{}\,\Phi_{t_0}^{t_1}\,\rho_S(t_0)\right]$ holds exactly 
as all other terms in Eq.~\eqref{eq:multi} vanish.
This equality is often referred to as the QRT,
even though it is a consequence of the actual QRT for
master equations of the GKSL form
(see discussion in Chap. 5 of Ref.~\cite{gard}).
In this particular case the knowledge of 
the map $\Phi$ and the initial state of the 
system is enough to fully characterize the system's dynamics. By extension we propose to define the {\it Markov regime}  as the range of parameters for which
\begin{equation}
\label{eq:definition}
 \langle\opt{o}{t_n}{n}{}\cdots\opt{o}{t_1}{1}{}\rangle=\text{Tr}_S\left[\op{o}{n}{}\,\Phi_{t_{n-1}}^{t_n}\cdots\op{o}{1}{}\Phi_{t_0}^{t_1}\,\rho_S(t_0)\right],
\end{equation}
for any number of operators $\op{o}{}{}$ acting on the system's Hilbert space
and any set of times $t_n\ge\cdots\ge t_1\ge t_0$. The proposed definition is universal and holds in both, classical and quantum domain. If the dynamics of the $n$-points correlation function of the reduced quantum system is given by the Eq. (\ref{eq:definition}) then the system is in the Markov regime of the system-environment interaction, otherwise the dynamics is non-Markovian. 

To interpret the above definition and to determine when a systems enters a non-Markovian regime according to it, we recall that
the effect of $\mathcal{P}$ is to `separate' 
whatever it acts on, which allows to reinterpret the object $\mathcal{Q}[\op{O}{}{}]$ as the carrier of the correlations
between the system and the environment ($\mathcal{P}+\mathcal{Q}=\mathcal{I}$).
To understand what this means let us explicitly interpret the term $\text{Tr}_S\left[\op{o}{2}{}\,\text{Tr}_E\left[\mathcal{P}\,S_{t_1}^{t_2}\,\op{o}{1}{}\,\mathcal{Q}\,S_{t_0}^{t_1}\,\mathcal{P}\opt{\rho}{t_0}{}{}\;\right]\right]$ (Eq.~\eqref{eq:mark}).
Here the system and environment are initially in the uncorrelated state $\mathcal{P}\rho(t_0)=
\opt{\rho}{t_0}{S}\otimes\op{\rho}{E}{}$ and undergo a joint
(free) evolution from time $t_0$ up to $t_1$. During this time interval correlations between
system and environment are generated
and the functional $\mathcal{Q}$ selects the part of the state containing these
correlations.
After this selection at $t=t_1$, the measurement of $\op{o}{1}{}$ is performed and
the system is let to evolve freely from time $t_1$ to time $t_2$.
At $t=t_2$ the total state is again separated by the action of the functional
$\mathcal{P}$ and a final measurement of the second observable $\op{o}{2}{}$ takes place.
Thus this term contributes to the total correlation function 
$\langle\opt{o}{t_2}{2}{}\opt{o}{t_1}{1}{}\rangle$
with the correlations developed during the free evolution
during the time interval $[t_0,t_1]$.
The role of possible initial correlations in
the dynamics of the system are taken
into account by the term 
$\text{Tr}_S\left[\op{o}{2}{}\,\text{Tr}_E\left[\mathcal{P}\,S_{t_1}^{t_2}\,\op{o}{1}{}\,\mathcal{P}\,S_{t_0}^{t_1}\,\mathcal{Q}\opt{\rho}{t_0}{}{}\;\right]\right]$.

If the application of $\mathcal{Q}$  
gives a non-zero value, the system and environment are correlated initially or at any point during the evolution. Using the analogy with the classical case, the definition of Markov regime for quantum dynamics 
given above can thus be reinterpreted as the 
dynamical regime in which system-environment correlations are negligible. This in turn allows us to write a compact 
expression for the multivariate correlation functions, namely the one given in Eq.~\eqref{eq:definition}.
Let us note that the role of correlations between system and environment has recently been pointed
out elsewhere in relation to the study of degree of non-Markovianity of a quantum 
map~\cite{laura}.

The definition given above follows naturally from the classical definition of Markovian systems.
This fact was already pointed out more than thirty years ago by Grishanin~\cite{grishanin1, grishanin2} for the
specific case of an atom interacting with a strong electromagnetic field.
Moreover, our formal expression naturally generalizes one
proposed by Lindblad himself~\cite{lind}.
However, our treatment and derivation above gives it a clear meaning in terms of correlations
generated in the joint system-environment dynamics.

{\it Examples }- 
In the following we will consider three exactly solvable toy models in order to
test the proposed definition of Markovianity: the spontaneous emission of a two level atom in 
vacuum, decoherence induced on a two level atom by a bosonic bath
and decoherence on the photon's polarization induced by interaction with its 
spatial degrees of freedom. Due to the solvability of these models we obtain the explicit form of the exact unitary operator describing the system-environment interaction and the exact master equation for the reduced system.We will then compare two point correlation functions
$\langle\opt{o}{t_2}{2}{}\opt{o}{t_1}{1}{}\rangle_\text{EXP}=\text{Tr}_S\left[\opt{o}{t_2}{2}{}\opt{o}{t}{1}{}\opt{\rho}{t_0}{}{}\right]$
calculated with the help of the unitary operator describing system-environment interaction, with  those calculated with the dynamical map $\Phi$ using Eq.~(\ref{eq:definition}), i.e.
$\langle\opt{o}{t_2}{2}{}\opt{o}{t_1}{1}{}\rangle_\text{M}=\text{Tr}_S\left[\op{o}{2}{}\,\Phi_{t_1}^{t_2}\,\op{o}{1}{}\,\Phi_{t_0}^{t_1}\,\rho_S(t_0)\right]$ \cite{open}.
The subscript `EXP' and `M' stand for `experimental' and `Markov', respectively, since the former 
describe a correlation function that would be measured in a real experiment, whereas
the latter are those calculated 
from the knowledge of the dynamical map $\Phi$.
Any difference between the two will be a signature of a transition 
from the Markov to the non-Markov regime.

The free Hamiltonian operators for the two-level atom and the field are given by
\begin{equation}
\op{H}{0}{a}=\frac{\omega_0}{2}\op{\sigma}{z}{}\quad\text{and}\quad
\op{H}{0}{f}=\sum_{{\bf k}}\omega_{{\bf k}}\,\op{b}{{\bf k}}{\dag}\op{b}{{\bf k}}{},
\end{equation}
and the interaction Hamiltonians, which describe the situations of decay and of decoherence are given by
\begin{equation}
\label{eq:ham1}
\op{H}{I}{(\text{decay})}=\op{\sigma}{+}{}\;\sum_{{\bf k}}g_{{\bf k}}\,\op{b}{{\bf k}}{}+\op{\sigma}{-}{}\;\sum_{{\bf k}}g_{{\bf k}}^{*}\,\op{b}{{\bf k}}{\dag},
\end{equation}
\begin{equation}
\label{eq:ham2}
\op{H}{I}{(\text{decoherence})}=\op{\sigma}{z}{}\;\sum_{{\bf k}}\left(g_{{\bf k}}\,\op{b}{{\bf k}}{}+g_{{\bf k}}^{*}\,\op{b}{{\bf k}}{\dag}\right).
\end{equation}
and 
\begin{equation}
\label{eq:ham3}
\op{H}{I}{(\text{engineered})}=\frac{\Delta n}{2}\op{\sigma^z}{}{}\;\sum_{{\bf k}}\op{b}{{\bf k}}{\dag}\op{b}{{\bf k}}{}.
\end{equation}

For both models the relative change $\epsilon$ 
\begin{equation}
\label{eq:epsilon}
\epsilon=1-\langle\opt{o}{t_2}{2}{}\opt{o}{t_1}{1}{}\rangle_\text{M}/\langle\opt{o}{t_2}{2}{}\opt{o}{t_1}{1}{}\rangle_\text{EXP}
\end{equation}
between the exact two point correlation functions and those calculated 
by assuming that the system is in a Markov regime is analyzed. 
For the spontaneous emission model the operators $\op{o}{1}{}$ and $\op{o}{2}{}$ are chosen to be $\op{o}{2}{}=\op{\sigma}{}{+}$ and $\op{o}{1}{}=\op{\sigma}{}{-}$
and the bosonic environment  is characterised by a {\it continuum} of electromagnetic field modes with a Lorentzian
spectral density $J(\omega)=(2\pi)^{-1}\gamma_0\lambda^2((\omega-\omega_0+\Delta)^{2}+\lambda^2)^{-1}$. Here, $\gamma_0$ is the coupling strength and the atom is initially in the excited level.

\begin{figure}[t]
\includegraphics[width=\linewidth]{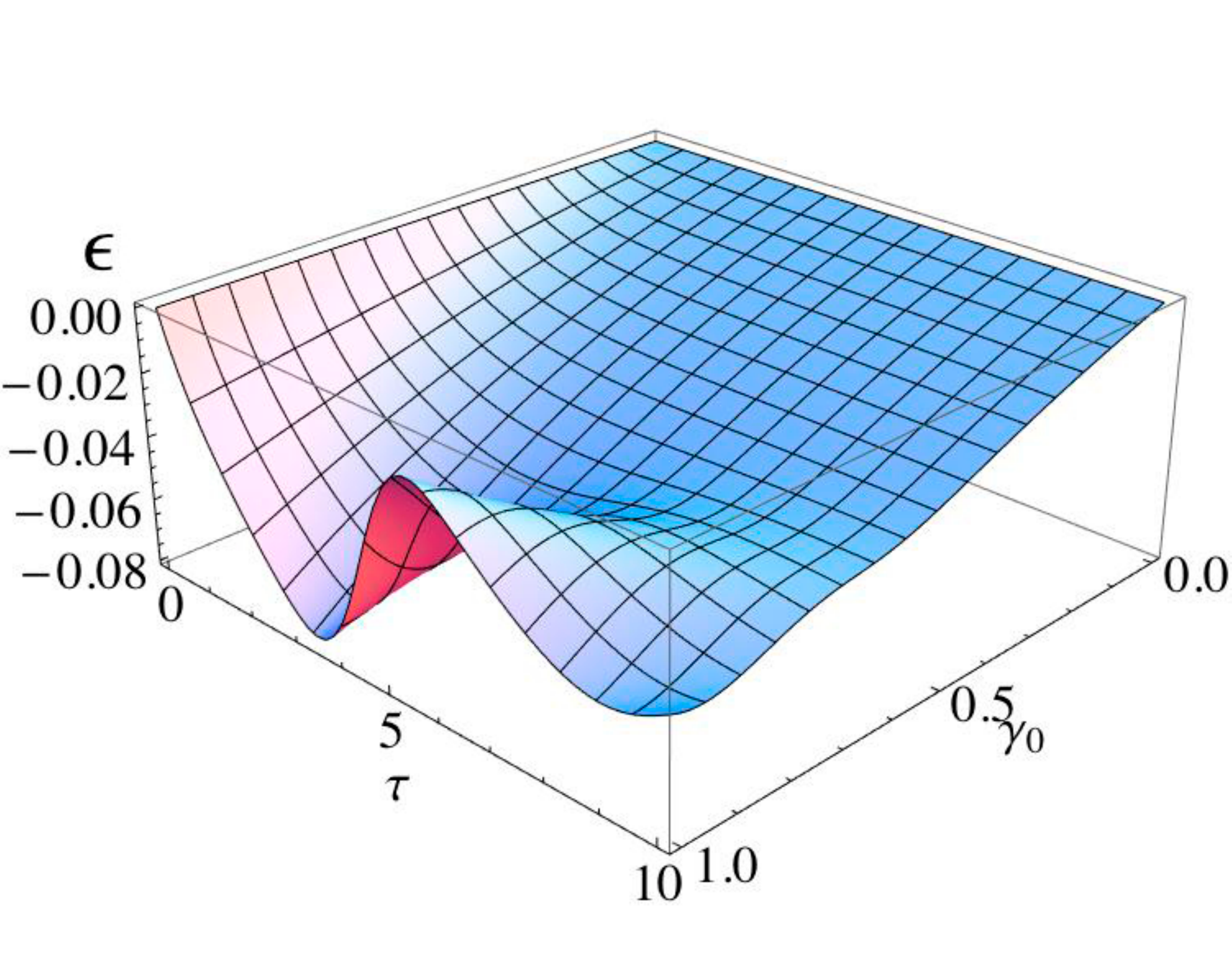}\\
\caption{(Color online) Relative change (Eq.~\ref{eq:epsilon}) of the correlation function
$\langle\opt{\sigma}{0.1+\tau}{+}{}\opt{\sigma}{0.1}{-}{}\rangle$ for the case of the spontaneous emission of a two level atom initially in its excited state and interacting with a bosonic bath. The 
spectral density of the bath is given by $J(\omega)=(2\pi)^{-1}\gamma_0\lambda^2((\omega-\omega_0+\Delta)^{2}+\lambda^2)^{-1}$ and it is assumed to be initially in the vacuum state. The parameter used are 
$\omega_0=20,\; \lambda=1.1,\; \Delta=0.2$.}
\label{fig:diss}
\end{figure}

Fig.~\ref{fig:diss} shows the dynamics of the relative change $\epsilon$ for system-bath coupling strength $\gamma_0$ between $0$ and $1$. 
As it is expected for small values of the coupling strength $\gamma_0$ the system is well in the Markov regime $\left(\langle\cdot\rangle_\text{M}\approx \langle\cdot\rangle_\text{EXP}\right)$. However, increasing the coupling strength $\gamma_0$ 
one can see that the dynamics of the system becomes non-Markovian and knowledge of the dynamical map $\Phi$ is not sufficient for the full description of the system.
It is important to mention that the correlation function 
$\langle \opt{\sigma}{t+\tau}{+}{}\opt{\sigma}{t}{-}{}\rangle$ is related to the (non-stationary) spectrum of the atom, which means that this behaviour is measurable~\cite{renaud77}. This is especially interesting since we can compare our result with recently suggested measures of non-Markovianity based on the trace distance \cite{breuer}, divisibility of the dynamical map \cite{div1,div2} and the volume of states accessible under the dynamical map \cite{volmap}. All these
measures are based on the analysis of the dynamical map only and predict the Markovian behaviour for the set of parameters used in Fig.~\ref{fig:diss}. However, as it is shown the residual information in the dynamical map is not enough to describe the dynamics of the reduced system, which implies non-Markovian dynamics.

For the second example describing pure decoherence of a two-state system interacting with a bosonic bath, whose 
spectral density is
$J(\omega)=(2\pi)^{-1}\gamma_0\lambda^2\omega(\omega^{2}+\lambda^2)^{-1}$, we find that $\langle\cdot\rangle_\text{M}= \langle\cdot\rangle_\text{EXP}$ at any time. This implies that decoherence is always a Markovian process in the case of a bosonic bath with a continuum of states and dynamical maps give  the full description of the system. However, measures based on the analysis of the properties of the dynamical maps \cite{breuer, div1,div2, volmap} predict non-Markovian evolution for some values of the parameters. The details of the calculations for both examples can be found in the Supplementary Material \cite{supp}.

\begin{figure}[t]
\includegraphics[width=9cm]{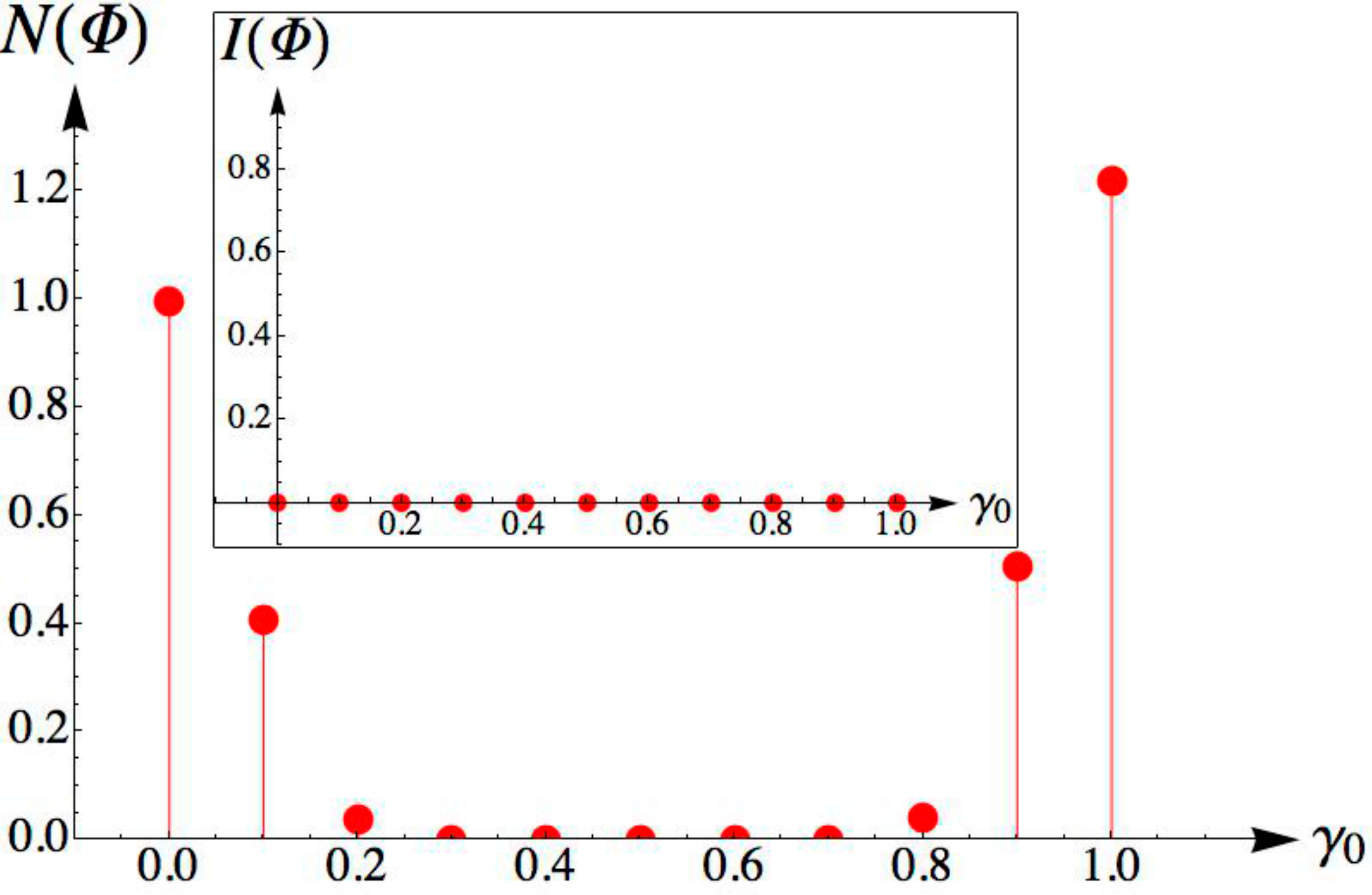}
\caption{(Color online). Values of the trace distance measure $N(\Phi)$ for the system described by 
Eq. (\ref{eq:ham3}) as a function of the control parameter $\gamma_0$. The two crossings from non-Markovian to Markovian regimes are clearly visible
and they correspond to the change of $|f(\omega)|^{2}$ from a double to a single peaked distribution.
In the inset we show the divisibility measure for the dynamical map. According to this measure the map is always Markovian.}
\label{fig:NMDiv}
\end{figure}

In Fig.~\ref{fig:NMDiv} we show values of the trace distance measure $N(\Phi)$ ~\cite{breuer, expBr, supp} for different values of $\gamma_0$, where $\Phi$ denotes the dynamical map describing the evolution of the sub-system of interest. We can clearly see, as already shown in Ref.~\cite{expBr}, that according to the trace distance measure  $N(\Phi)$
the systems goes from a non-Markovian to a Markovian regime and then back to the non-Markovian one.  These crossings correspond 
to the crossing of $|f(\omega)|^{2}$ from a double to a single peaked distribution (see Supplementing Material).
Nevertheless, we found that two point correlation functions calculated with the dynamical map always
coincide with the exact ones so that $\epsilon=0$ even for this decoherence process.

For all the above cases we also checked the so-called divisibility measure $I(\Phi)$ ~\cite{div1,div2} and we found that 
it is zero in all of them (see inset in Fig.~\ref{fig:NMDiv}).
This is not surprising since it can be shown that the maps arising from the total unitary evolution
given by the Hamiltonians in Eqs. (\ref{eq:ham1}), (\ref{eq:ham2}), (\ref{eq:ham3}) are divisible ones.
We would like to stress two points. First of all the decay case we considered offers a perfect example of a quantum system for which it is possible
to find a dynamical map with the divisibility property but whose two (and higer) order correlation
functions have a non trivial dependence upon the one point ones (mean values). This is remarkable
because this was exactly where we started from in order to define Markov regimes in analogy
with the classical case~\cite{hanggi1,hanggi2}.
Second this is another example of the discrepancy between the two most used measures of non-Markovianity~\cite{comp1,comp2}.

{\it Fluctuation relations and Markov regime} -
Over the past thirty years the study of fluctuation relations
for closed quantum systems in analogy to the classical case has been an active research area. 
Different approaches have been pursued, depending on the nature of the system under study
(see Ref.~\cite{FRLindblad} for brief review).
At the heart of the problem in the quantum case is the difficulty of properly defining a quantity, 
which satisfies a fluctuation relation while at the same time representing the work done on the system.
For an open quantum system the problem is even more complicated, because one has to
identify the work done on the environment (heat) as well.
An approach that can treat this problem successfully is to use an accompanying reference dynamics of
the stationary state.

In the following we want to show that our definition of the Markovian regime 
allows for the possibility of defining a super operator for which 
a fluctuation relation holds. We will follow the discussion in Ref.~\cite{FRLindblad}
and extend it to the case where the dynamics is not described by a Lindlbad super operator~\cite{nonGKSL}.

To this end we consider the Liouvillian $\mathcal{L}[\circ]=-\imath[\opt{H}{t}{}{},\circ]$, where the total Hamiltonian is in general
time-dependent (external driving).
The instantaneous stationary state $\opt{\pi}{t}{}{}$ of the system is defined by means of
the relations
$\opt{\pi}{t}{}{}=\text{Tr}_E[\opt{\tilde \rho}{t}{\pi}{}]$ and
$\partial\opt{\pi}{t}{}{}/\partial t=\text{Tr}_E[\mathcal{L}[\opt{\tilde \rho}{t}{\pi}{}]]=0$. 
The state $\opt{\pi}{t}{}{}$ is the accompanying state.
Such a stationary state is generally time dependent and we shall assume that
its equation of motion is given by $\partial\opt{\pi}{t}{}{}/\partial t = -\mathcal{W}(t)\opt{\pi}{t}{}{}$.
Following the discussion in Ref.~\cite{FRLindblad} and by means of the decomposition 
of the correlation functions as in Eq.~(\ref{eq:definition})
it is possible to show that for the mean value of any observable of the system
one has
\begin{equation}
\label{eq:FR}
\text{Tr}_S[\opt{\pi}{t}{}{}\opt{o}{t_0}{}{}]=\text{Tr}_S[\opt{\pi}{t_0}{}{}e^{-\int_{t_0}^{t}d\tau\;\mathcal{W}(\tau)}\opt{o}{t}{}{}]-\mathcal{R(\mathcal{Q})}.
\end{equation}
In the case of a GKSL form $\mathcal{R(\mathcal{Q})}=0$ and we recover the case discussed in Ref.~\cite{FRLindblad}. 
But $\mathcal{R(\mathcal{Q})}=0$ would also hold for a system whose dynamics is not driven by
a GKSL master equation but for which Eq.~(\ref{eq:definition}) holds for any $n$.
This result shows that it is possible to
derive a fluctuation relation for the variable $\mathcal{W}$ in the Markov regime.
Furthermore, in the non-Markovian regime we have $\mathcal{R(\mathcal{Q})}\ne0$,
due to the presence of initial/dynamical correlations which are not negligible over
the time scale considered.

{\it Conclusions } -
We have presented a universal definition of non-Markov regime that holds in the quantum and the classical domain.
Only in the Markov case the knowledge of both, the density operator's 
propagator (dynamical map) and the initial state of the system, is enough for a {\it full} description
of the system dynamics, {\it i.e.} for the evaluation of any $n$-points (in time) correlation functions. In the case of the quantum non-Markovian  dynamics the knowledge of the dynamical map and the initial state of the system does not allow to fully describe the reduced system dynamics. Spontaneous emission of a two-level system exemplified our definition of the Markov regime, by explicitly demonstrating that the knowledge of the exact dynamical map and the initial condition is not sufficient to describe the $2$-point correlation function of the system. 
The transition into a non-Markov regime is induced by the build up
of correlations between system and environment, which in turn are propagated during the 
joint evolution. Once this has happened, as in the classical case~\cite{hanggi1,hanggi2}, 
 the $n$-point correlation functions
need to be calculated by means of $n$-points maps 
$\Phi_{t_n,\;\cdots\;,t_1,t_0}^{(n)}$ which, in general, are
not trivially related to $\Phi_{t_0}^{t}$. Finally, we have shown that in the Markov regime it is possible to define a ``work functional'' in analogy to 
the one introduced in Ref.~\cite{FRLindblad} for the case of GKSL master equation
for which a fluctuation relation holds.

\section{Acknowledgements}
This work was supported by Science Foundation Ireland 
under grant numbers 10/IN.1/I2979 and 10/IN.1/I2979 - STTF11, the Irish Council for Science, Engineering and Technology 
through the Embark Initiative (RS/2000/137). 
This work is based upon research supported by the South African Research 
Chair Initiative of the Department of Science and Technology and National Research 
Foundation. NLG is grateful to S. Lorenzo and J. Goold 
for useful discussions and thanks F. Plastina and K. M\o lmer for pointing
out the relation between this work and the quantum regression theorem in quantum optics.
NLG also thanks the NITheP at the UKZN for their hospitality.

\clearpage
\begin{widetext}

\begin{center}
\Large{Supplementing material}
\end{center}

\section{Decay of a two-level atom}
\label{sec:dec}
In this section we consider the free decay of a two level atom coupled to a bosonic
bath in its vacuum state.
The Hamiltonian operator is given by 

\begin{equation}
\begin{aligned}
\op{H}{}{}&=\op{H}{0}{}+\op{H}{I}{},\\
\op{H}{0}{}&=\frac{\omega_0}{2}\op{\sigma}{z}{}+\sum_{{\bf k}}\omega_{{\bf k}}\,\op{b}{{\bf k}}{\dag}\op{b}{{\bf k}}{},\\
\op{H}{I}{}&=\op{\sigma}{+}{}\;\sum_{{\bf k}}g_{{\bf k}}\,\op{b}{{\bf k}}{}+\op{\sigma}{-}{}\;\sum_{{\bf k}}g_{{\bf k}}^{*}\,\op{b}{{\bf k}}{\dag}.\\
\end{aligned}
\end{equation}

\subsection{Evolution operator}
It is in general not possible to find a closed formula for the evolution operator.
Nevertheless we restrict to the single excitation case thus getting:

\begin{equation}
\begin{aligned}
\opt{U}{t}{}{}=e^{-\imath \op{H}{}{} t}&=c_{0}(t) \ketbra{0,0}{0,0}+c_{1}(t)\ketbra{1,0}{1,0}
+\sum_{{\bf q}} c_{{\bf q}} \ketbra{0,1_{{\bf q}} }{0,1_{{\bf q}}}\nonumber\\
&+\sum_{{\bf q}}\sum_{{\bf p}\neq{\bf q}} c_{{\bf q}{\bf p}} \ketbra{0,1_{{\bf q}} }{0,1_{{\bf p}}}+\sum_{{\bf q}} \lambda_{{\bf q}} \ketbra{0,1_{{\bf q}} }{1,0}+\sum_{{\bf q}} \mu_{{\bf q}} \ketbra{1,0}{0,1_{{\bf q}}}.
\end{aligned}
\end{equation}

Since $[\opt{U}{t}{}{},\op{H}{}{}]=0$ the following equalities hold to 
be true:

\begin{equation}
\begin{aligned}
\mu_{{\bf q}}(t)&=\frac{g_{{\bf q}}}{g_{{\bf q}}^*}\,\lambda_{{\bf q}}(t),\\
c_{{\bf q}}(t)&=c_1(t)+\left(\frac{\omega_{{\bf q}}-\omega_{0}}{g_{{\bf q}}^{*}}-\sum_{{{\bf p}}\neq{{\bf q}}}\frac{g_{{\bf p}}^{*}}{g_{{\bf q}}^{*}}\frac{g_{{\bf p}}-g_{{\bf q}}}{\omega_{{\bf q}}-\omega_{{\bf p}}}\right)\lambda_{{\bf q}}(t),\nonumber\\
c_{{\bf q}{\bf p}}(t)&=\frac{g_{{\bf p}}-g_{{\bf q}}}{\omega_{{\bf q}}-\omega_{{\bf p}}}\lambda_{{\bf q}}(t),
\end{aligned}
\end{equation}
so that one has to solve only for the independent variables, namely $c_1(t)$ and $\lambda_{{\bf q}}^{*}(t)$.
By using the equation $d\opt{ U}{t}{}{}/dt=-\imath \op{H}{}{}\opt{U}{t}{}{}$
and by mean of a change of variables $\tilde c_1(t)=c_1(t) e^{\imath \frac{\omega_0}{2} t}$ and $\tilde \lambda_{{\bf q}}(t)=\lambda_{{\bf q}}(t) e^{\imath \frac{\omega_0}{2} t}$ 
one is led to solve the following equations:

\begin{equation}
\begin{aligned}
\frac{d}{dt}c_0(t)&=\imath\, \omega_0\, c_0(t),\\
\frac{d}{dt}\tilde c_1(t)&=-\imath \,\sum_{{\bf q}} g_{{\bf q}} \,\tilde \lambda_{{\bf q}}(t),\\
\frac{d}{dt}\tilde \lambda_{{\bf q}}(t)&=-\imath\, (\omega_{{\bf q}}-\omega_0)\,\tilde \lambda_{{\bf q}}(t)-\imath\, g_{{\bf q}}^{*} \,\tilde c_1(t),\\
\end{aligned}
\end{equation}
which are to be solved with the initial conditions
$c_0(0)=1,\;\tilde c_1(0)=1,\;\tilde \lambda_{{\bf q}}(0)=0$.
The solution to the first equation is given by $c_0(t)=e^{\imath \frac{\omega_0}{2} t}$.
The solution for the last two is obtained by means of the Laplace transform as:

\begin{equation}
\begin{aligned}
\mathcal{L}[\tilde c_1(t)](s)&=\left(s+\sum_{{\bf q}}\frac{|g_{{\bf q}}|^2}{s+\imath(\omega_{{\bf q}}-\omega_0)}\right)^{-1},\\
\mathcal{L}[\tilde \lambda_{{\bf q}}(t)](s)&=-\imath \frac{g_{{\bf q}}(t)^{*}}{s+\imath (\omega_{{\bf q}}-\omega_0)} \mathcal{L}[\tilde c_1(t)](s).\\
\end{aligned}
\end{equation}
By taking the continuum limit  we have

\begin{equation}
\sum_{{\bf q}}\frac{|g_{{\bf q}}|^2}{s+\imath(\omega_{{\bf q}}-\omega_0)} \rightarrow \int_0^{\infty}d\omega \,\frac{J(\omega)}{s+\imath(\omega-\omega_0)},
\end{equation}
where we introduced the ``spectral density'' of the bath $J(\omega)$. We next note that 

\begin{equation}
\int_0^{\infty}d\omega \frac{J(\omega)}{s+\imath(\omega-\omega_0)}=\mathcal{L}\left[f(t)\right](s),
\end{equation}
where $f(t)=\int_0^{\infty}d\omega J(\omega)e^{-\imath(\omega-\omega_0)t}$.
The above equations then become:

\begin{equation}
\label{solU}
\begin{aligned}
\mathcal{L}[\tilde c_1(t)](s)&=\frac{1}{\left(s+\mathcal{L}[f(t)](s)\right)},\\
\mathcal{L}[\tilde \lambda_{{\bf q}}(t)](s)&=-\imath \frac{g_{{\bf q}}(t)^{*}}{s+\imath (\omega_{{\bf q}}-\omega_0)} \mathcal{L}[\tilde c_1(t)](s).\\
\end{aligned}
\end{equation}

Once the spectral density has been chosen then it is possible to invert 
the two above equations to find the exact evolution operator
in the zero and one excitation sector of the Hilbert space.

\subsection{Density matrix propagator}

Given now a system described by the same Hamiltonian as in the previous section,
addressing the case of one excitation approximation the total state of the system
can be written as $a_0(t)\ket{0,0}+a_1(t)\ket{1,0}+\sum_{{\bf q}}a_{{\bf q}}(t)\ket{0,1_{{\bf q}}}$
It is possible to show that in the one excitation approximation and for the 
bath in the vacuum state the master equation governing the dynamics of the system
is given by:

\begin{equation}
\frac{d}{dt}\opt{\rho}{t}{S}{}=-\frac{\imath }{2}\left(\omega_0+\frac{S(t)}{2}\right)[\op{\sigma}{z}{},\opt{\rho}{t}{S}{}]
+\gamma(t)\left(\op{\sigma}{-}{}\opt{\rho}{t}{S}{}\op{\sigma}{+}{}-\frac{1}{2}\left\{\op{\sigma}{+}{}\op{\sigma}{-}{},\opt{\rho}{t}{S}{}\right\}\right),\nonumber
\end{equation}
where $\gamma(t)+\imath S(t)= 2 \int_0^{t} d\tau\; f(t-\tau)\, G(\tau,0)/G(t,0)$, 
$f(t)=\int_0^{\infty}d\omega J(\omega)e^{-\imath(\omega-\omega_0)t}$ and $G(t,0)$
such that $d G(t,0)/dt=-\int_0^{t} d\tau\; f(t-\tau)\, G(\tau,0)$. 

The ``propagator'' is then given by $\Phi_{0}^{t}=e^{\int_{0}^{t}d\tau\; \mathcal{L}(\tau)}$.
Since the coefficient $\gamma(t)$ can take on negative values then for certain
instants in time the super-operator $\mathcal{L}(t)$ could be not of the Lindbland form.
In what follows we will have to calculate the action of the super-operator on operators 
acting on the Hilbert space of the system.
In order to simplify this calculations we will use the damping basis, i.e. the basis made 
up of operators $\opt{\Lambda}{t}{i}{}$ such that $\mathcal{L}(t)[\op{\Lambda}{}{i}]=\lambda_i(t)\, \opt{\Lambda}{t}{i}{}$. In fact the ``eigen-operators'' $\op{\Lambda}{i}{}$
are time independent: 
\begin{center}
\begin{tabular}{lcl}
 & &\\
$\op{\Lambda}{0}{}=\frac{1}{2}(\op{1}{}{}-\op{\sigma}{z}{}),$& &$\lambda_0(t)=0,$ \\ 
$\op{\Lambda}{1}{}=\op{\sigma}{+}{},$ & &$\lambda_1(t)=-\imath\,\left(\omega_0+\frac{S(t)}{2}\right)-\frac{\gamma(t)}{2},$\\
$\op{\Lambda}{2}{}=\op{\sigma}{-}{},$ & &$\lambda_2(t)=\imath\,\left(\omega_0+\frac{S(t)}{2}\right)-\frac{\gamma(t)}{2},$\\
$\op{\Lambda}{3}{}=\op{\sigma}{z}{},$& &$\lambda_3(t)=-\gamma(t).$\\
\end{tabular}
\end{center}
We can thus write any operator as $\op{o}{}{}=\sum_{i} c^{i}\, \op{\Lambda}{i}{}$
with $c^{i}=\mathrm{Tr}(\check{\Lambda}^{i}\op{o}{}{})$ where the dual of the damping 
basis 's elements are such that $\mathrm{Tr}(\check{\Lambda}^{i}\op{\Lambda}{j}{})=\delta_{j}^{i}$.

Let now define the matrices $(A_{\alpha})_{i}^{j}=\mathrm{Tr}\big[\check{\Lambda}^{j}\op{\sigma}{\alpha}{}\op{\Lambda}{i}{}\big]$ with help of which it is easy to calculate multivariate correlation functions.
The two-point ones are 

\begin{equation}
\begin{aligned}
<\;\opt{\sigma}{t+\tau}{\alpha}{}\opt{\sigma}{t}{\beta}{}\;>_P=&&\mathrm{Tr}\bigg[\opt{\sigma}{0}{\alpha}{}\Phi_{0}^{\tau}\opt{\sigma}{0}{\beta}{}\Phi_{0}^{t}\opt{\rho}{0}{S}{}\bigg]\\
=&&\sum_{i,j}(A_{\alpha})_{j}^{0}(A_{\beta})_{i}^{j} e^{L_j(\tau)} e^{L_i(t)}c^{i},
\end{aligned}
\end{equation}
where $L_i(t)=\int_0^{t}ds\; \lambda_i(s)$.

\subsection{Exact two-point correlation functions}

In order to calculate the multivariate correlation functions between 
$n$ operators $\op{o}{}{}$ acting on the system's Hilbert space we need to calculate
$\opt{o}{t}{}{}=\opt{U}{t}{}{\dag}\op{o}{}{}\opt{U}{t}{}{}$.
From the knowledge of the evolution this is an easy task Eq. (\ref{solU}).

\section{Decoherence of a two-level atom in a thermal bath}
\label{sec:deco}
In this section we consider the decoherence of a qubit in vacuum.
The Hamiltonian operator is given by 

\begin{equation}
\begin{aligned}
\op{H}{}{}&=\op{H}{0}{}+\op{H}{I}{},\\
\op{H}{0}{}&=\frac{\omega_0}{2}\op{\sigma}{z}{}+\sum_{{\bf k}}\omega_{{\bf k}}\,\op{b}{{\bf k}}{\dag}\op{b}{{\bf k}}{},\\
\op{H}{I}{}&=\op{\sigma}{z}{}\;\sum_{{\bf k}}\left(g_{{\bf k}}^{*}\,\op{b}{{\bf k}}{}+g_{{\bf k}}\,\op{b}{{\bf k}}{\dag}\right).\\
\end{aligned}
\end{equation}

\subsection{Evolution operator}
In the interaction picture the (total) time evolution operator is 
\begin{equation}
\opt{U}{t}{}{}=e^{-\imath\int_0^{t}d \tau \opt{H}{\tau}{I}{} }=\op{1}{S}{}\otimes \cosh(\op{P}{}{})+\op{\sigma}{z}{}\otimes \sinh(\op{P}{}{}),
\end{equation}
where $\op{P}{}{}=\sum_{{\bf k}}\left(\alpha_{{\bf k}}\,\op{b}{{\bf k}}{\dag}-\alpha_{{\bf k}}^{*}\,\op{b}{{\bf k}}{}\right)$ and 
$\alpha_{{\bf k}}=g_{{\bf k}} (1-e^{\imath \omega_{{\bf k}} t})\omega_{{\bf k}}^{-1}$.

\subsection{Density matrix propagator}
The density matrix propagator is, by definition, given by:

\begin{equation}
\begin{aligned}
\Phi_{0}^{t}[\opt{\rho}{0}{S}{}]&=\mathrm{Tr}_E\left[\opt{U}{t}{}{}\opt{\rho}{0}{}{}\opt{U}{t}{}{\dag}\right]\\
&=\opt{\rho}{0}{S}{}\;\mathrm{Tr}_E\left[\cosh(\op{P}{}{})\opt{\rho}{0}{E}{}\cosh(\op{P}{}{})\right]-\op{\sigma}{z}{}\opt{\rho}{0}{S}{}\op{\sigma}{z}{}\;\mathrm{Tr}_E\left[\sinh(\op{P}{}{})\opt{\rho}{0}{E}{}\sinh(\op{P}{}{})\right]\\
&+\op{\sigma}{z}{}\opt{\rho}{0}{S}{}\;\mathrm{Tr}_E\left[\sinh(\op{P}{}{})\opt{\rho}{0}{E}{}\cosh(\op{P}{}{})\right]-\opt{\rho}{0}{S}{}\op{\sigma}{z}{}\;\mathrm{Tr}_E\left[\cosh(\op{P}{}{})\opt{\rho}{0}{E}{}\sinh(\op{P}{}{})\right],
\end{aligned}
\end{equation}
where as usual $\opt{\rho}{0}{S}{}=\mathrm{Tr}_E\left[\opt{\rho}{0}{}{}\right]$. We now assume that 
the environment is initially in a thermal state $\opt{\rho}{0}{E}{}=e^{-\beta\sum_{{\bf k}}\omega_{{\bf k}}\,\op{b}{{\bf k}}{\dag}\op{b}{{\bf k}}{}}/Z$.
It is thus possible to perform the traces over the environment's Hilbert space by switching to the phase space
and noticing that we will have to calculate terms such as $\mathrm{Tr}_E\left[e^{\xi_{1} \op{P}{}{}}\opt{\rho}{0}{E}{}e^{\xi_{2} \op{P}{}{}}\right]=\prod_{\bf k}\mathrm{Tr}_{\bf k}\left[e^{\xi_{1} \op{P}{\bf k}{}}\opt{\rho}{0}{E}{}e^{\xi_{2} \op{P}{\bf k}{}}\right]$ where $\xi_{1,2}=\pm 1$ and $\op{P}{k}{}=\alpha_{{\bf k}}\,\op{b}{{\bf k}}{\dag}-\alpha_{{\bf k}}^{*}\,\op{b}{{\bf k}}{}$.
Therefore we only have to calculate the quantity $\mathrm{Tr}_{\bf k}\left[e^{\xi_{1} \op{P}{\bf k}{}}\opt{\rho}{0}{E}{}e^{\xi_{2} \op{P}{\bf k}{}}\right]$ for a single harmonic oscillator.
In order to do so, let us first define the displacement operator $\mathcal{D}(\xi \alpha_{\bf k})=e^{\xi \op{P}{\bf k}{}}$. 
Then by using the relation $\opt{D}{\alpha}{}{}\opt{D}{\beta}{}{}=e^{(\alpha\beta^{*}-\alpha^{*} \beta)/2}\opt{D}{\alpha+\beta}{}{}$ we can rewrite

\begin{equation}
\begin{aligned}
 &\mathrm{Tr}_{\bf k}\left[e^{\xi_{1} \op{P}{\bf k}{}}\opt{\rho}{0}{E}{}e^{\xi_{2} \op{P}{\bf k}{}}\right]=\mathrm{Tr}_{\bf k}\left[\opt{\rho}{0}{E}{}e^{\xi_{2} \op{P}{\bf k}{}}e^{\xi_{1} \op{P}{\bf k}{}}\right]\\
 &=\mathrm{Tr}_{\bf k}\left[\opt{\rho}{0}{E}{}\opt{D}{\xi_{2}\alpha_{\bf k}}{}{}\opt{D}{\xi_{1}\alpha_{\bf k}}{}{}\right]=\mathrm{Tr}_{\bf k}\left[\opt{\rho}{0}{E}{}\opt{D}{(\xi_{1}+\xi_{2})\alpha_{\bf k}}{}{}\right]\\
 &=e^{\frac{(\xi_{1}+\xi_{2})^2}{2}|\alpha_{\bf k}|^2}\mathrm{Tr}_{\bf k}\left[\opt{\rho}{0}{E}{}e^{(\xi_{1}+\xi_{2})\alpha_{\bf k}\op{b}{\bf k}{\dag}}e^{-(\xi_{1}+\xi_{2})\alpha_{\bf k}^{*}\op{b}{\bf k}{}}\right]=e^{\frac{(\xi_{1}+\xi_{2})^2}{2}|\alpha_{\bf k}|^2}\chi_{\bf k}((\xi_{1}+\xi_{2})\alpha_{\bf k},(\xi_{1}+\xi_{2})\alpha_{\bf k}).
\end{aligned}
\end{equation}

The function $\chi_{\bf k}(\alpha_{\bf k},\alpha_{\bf k})$ is called the characteristic
function of the harmonic oscillator since its knowledge allows for the calculation of 
any product $\langle\op{b}{\bf k}{\dag s}\op{b}{\bf k}{r}\rangle$.
For a thermal state we have $\chi_{\bf k}((\xi_{1}+\xi_{2})\alpha_{\bf k},(\xi_{1}+\xi_{2})\alpha_{\bf k})=e^{-(\xi_{1}+\xi_{2})^2|\alpha_{\bf k}|^2/(e^{\beta\omega_{\bf k}}-1)}$.
It is now easy to calculate the terms entering the density matrix propagator which are given by:

\begin{equation}
 \begin{aligned}
  \mathrm{Tr}_E\left[\cosh(\op{P}{}{})\opt{\rho}{0}{E}{}\cosh(\op{P}{}{})\right]&=\frac{1}{2}\left(\sum\limits_{\bf k}e^{-2 |\alpha_{\bf k}|^2\coth\left(\frac{\beta \omega_{\bf k}}{2}\right)}+1\right),\\
  \mathrm{Tr}_E\left[\sinh(\op{P}{}{})\opt{\rho}{0}{E}{}\sinh(\op{P}{}{})\right]&=\frac{1}{2}\left(\sum\limits_{\bf k}e^{-2 |\alpha_{\bf k}|^2\coth\left(\frac{\beta \omega_{\bf k}}{2}\right)}-1\right),\\
  \mathrm{Tr}_E\left[\cosh(\op{P}{}{})\opt{\rho}{0}{E}{}\sinh(\op{P}{}{})\right]&=\mathrm{Tr}_E\left[\sinh(\op{P}{}{})\opt{\rho}{0}{E}{}\cosh(\op{P}{}{})\right]=0.
 \end{aligned}
\end{equation}

The density matrix propagator is thus given by:

\begin{equation}
\Phi_{0}^{t}\;:\; \opt{\rho}{0}{S}{}\rightarrow\frac{1}{2}\left((1+e^{-g(t)})\,\opt{\rho}{0}{S}{}+(1-e^{-g(t)})\,\op{\sigma}{z}{}\,\opt{\rho}{0}{S}{}\,\op{\sigma}{z}{}\right),
\end{equation}
where the continuum limit has been taken in order to define the function 
\begin{equation}
  4\sum_{{\bf k}}\frac{|g_{{\bf k}}|^2}{\omega_{{\bf k}}^2}(1-\cos(\omega_{{\bf k}} t)) \cosh\left(\frac{\beta \omega_{\bf k}}{2}\right)\rightarrow g(t)=4 \int_{0}^{\infty}d\omega\,\frac{J(\omega)}{\omega^2}(1-\cos(\omega t))\cosh\left(\frac{\beta \omega}{2}\right).
\end{equation}

By using the set of operators $\{\op{\sigma}{0}{}=\op{1}{S}{},\op{\sigma}{+}{},\op{\sigma}{-}{},\op{\sigma}{z}{}\}$ as the damping basis and their duals one can write $\opt{\rho}{t}{S}{}=\sum_{i}c^i(t)\,\op{\sigma}{i}{}$ where 
\begin{equation}
\begin{aligned}
c^i(t)&=\sum_{j}v_{j}^{i}(t)\, c^{j}(0),\nonumber\\ 
v_{j}^{i}(t)&=\frac{1+e^{-g(t)}}{2} \, \delta_{j}^{i}+\frac{1-e^{-g(t)}}{2} \,(A^{zz})_{j}^{j},\nonumber\\ 
(A^{zz})_{j}^{j}&=Tr\bigg[\opd{\sigma}{}{j}\,\op{\sigma}{z}{}\op{\sigma}{j}{}\op{\sigma}{z}{}\bigg].
\end{aligned}
\end{equation}
As in the previous section we have:

\begin{equation}
\begin{aligned}
<\;\opt{\sigma}{t+\tau}{\alpha}{}\opt{\sigma}{t}{\beta}{}\;>_P&=Tr\bigg[\opt{\sigma}{0}{\alpha}{}\Phi_{0}^{\tau}\opt{\sigma}{0}{\beta}{}\Phi_{0}^{t}\opt{\rho}{0}{S}{}\bigg]\nonumber\\
&= \sum_{k,l,j,i}(B_{\alpha})_{k}^{0}v_{l}^{k}(\tau)(B_{\beta})_{j}^{l} v_{i}^{j}(t)\, c^i(0),
\end{aligned}
\end{equation}
where $(B_{\alpha})_{i}^{j}=Tr\big[\check{\Lambda}^{j}\op{\sigma}{\alpha}{}\op{\Lambda}{i}{}\big]$.

\subsection{Exact two-point correlation functions}

In order to calculate the exact two-points correlation function of any two
operators $\op{o}{1}{}$ and $\op{o}{2}{}$ acting on the system's Hilbert 
space only we have to calculate $\opt{o}{t_2}{1}{}\opt{o}{t_2}{2}{}=\opt{U}{t_1}{}{\dag}\opt{o}{0}{1}{}\opt{U}{t_1}{}{}\opt{U}{t_2}{}{\dag}\opt{o}{0}{2}{}\opt{U}{t_2}{}{}$.
Assuming the bosonic environment is in a thermal state $\op{\rho}{E}{}=e^{-\beta\sum_{{\bf k}}\omega_{{\bf k}}\,\op{b}{{\bf k}}{\dag}\op{b}{{\bf k}}{}}/Z$ and after lengthy calculations we get:

\begin{equation}
\begin{aligned}
Tr_E\big[\opt{o}{t_1}{1}{}\opt{o}{t_2}{2}{}\,\op{\rho}{E}{}\big]&=f_{1}(t_1,t_2)\,\opt{o}{0}{1}{}\,\opt{o}{0}{2}{}+f_{2}(t_1,t_2)\,\op{\sigma}{z}{}\,\opt{o}{0}{1}{}\,\opt{o}{0}{2}{}\,\op{\sigma}{z}{}\\
&\nonumber\\
&+f_{3}(t_1,t_2)\,\op{o}{0}{1}{}\,\op{\sigma}{z}{}\,\op{o}{0}{2}{}\,\op{\sigma}{z}{}+f_{4}(t_1,t_2)\,\op{\sigma}{z}{}\,\op{o}{0}{1}{}\,\op{\sigma}{z}{}\,\op{o}{0}{2}{},\\
&\\
f_{1}(t_1,t_2)&=\frac{1}{4}\left(1+e^{-g(t_1)}+e^{-g(t_2)}+e^{-g(t_1)-g(t_2)+2h(t_1,t_2)}\right),\\
f_{2}(t_1,t_2)&=\frac{1}{4}\left(1-e^{-g(t_1)}-e^{-g(t_2)}+e^{-g(t_1)-g(t_2)+2h(t_1,t_2)}\right),\\
f_{3}(t_1,t_2)&=\frac{1}{4}\left(1+e^{-g(t_1)}-e^{-g(t_2)}-e^{-g(t_1)-g(t_2)+2h(t_1,t_2)}\right),\\
f_{4}(t_1,t_2)&=\frac{1}{4}\left(1-e^{-g(t_1)}+e^{-g(t_2)}-e^{-g(t_1)-g(t_2)+2h(t_1,t_2)}\right),\\
\end{aligned}
\end{equation}
where the continuum limit has been taken again to define the new function:

\begin{equation}
\begin{aligned}
&2\sum_{{\bf k}}\frac{|g_{{\bf k}}|^2}{\omega_{{\bf k}}^2}\left(1-e^{-\imath\omega_{{\bf k}} t_1}-e^{-\imath\omega_{{\bf k}} t_2}+e^{-\imath\omega_{{\bf k}} (t_1-t_2)}\right) \cosh\left(\frac{\beta \omega_{\bf k}}{2}\right)\rightarrow h(t_1,t_2),\\
 &h(t_1,t_2)=2 \int_{0}^{\infty}d\omega\,\frac{J(\omega)}{\omega^2}\left(1-e^{-\imath\omega t_1}-e^{-\imath\omega t_2}+e^{-\imath\omega(t_1-t_2)}\right)\cosh\left(\frac{\beta \omega}{2}\right).
 \end{aligned}
\end{equation}

\section{Engineered decoherence of a two-level atom with an optical setup}

\subsection{Evolution operator}

In Ref.~\cite{expBr} it has been shown that the unitary evolution of the system plus
environment is given by the unitary operator:

\begin{equation}
\begin{aligned}
 \opt{U}{t,t_0}{}{}&=e^{-\imath \frac{\Delta n}{2}\op{\sigma^z}{}{}\sum\limits_{k} \omega(k)\op{n}{k}{} (t-t_0)},\\
 \op{n}{k}{}&=\op{a}{k}{\dag}\op{a}{k}{},
\end{aligned}
\end{equation}
where $\Delta n=n_H-n_V$, and $n_H\; (n_V)$ are the refractive indeces for horizontal (vertical) polarized beams
(the photons go through a bi-refractive material), and $\op{a}{k}{}$ and $\op{a}{k}{\dag}$
are bosonic annihilation and creation operators for the mode $k$ with frequency $\omega(k)$.

The initial state of the total system has been chosen to be:

\begin{equation}
\begin{aligned}
 \ket{\Psi_{1,2}}&=\ket{\phi_{1,2}}\ket{\chi},\\
 \ket{\phi_{1,2}}&=\frac{1}{2}\left(\ket{H}\pm\ket{V}\right),\\
 \ket{\chi}&=\int d\omega\; f(\omega)\ket{1_\omega}.
\end{aligned}
\end{equation}

We considered an initial distribution of photons in the frequency domain which can be changed as a function
of a driving parameter that we will call $\gamma_0$.
The different distributions as a function of $\gamma_0\in [0,1]$ is shown in Fig.~\ref{fig:fofomega}.
The main feature to be noted here is that while changing $\gamma_0$ we go from a double picked
distribution to a single picked one and then back to a double picked one.

\begin{figure}[t]
\begin{center}
\begin{tabular}{ccc}
{\bf a)}&&{\bf b)}\\
&&\\
\includegraphics[width=8cm]{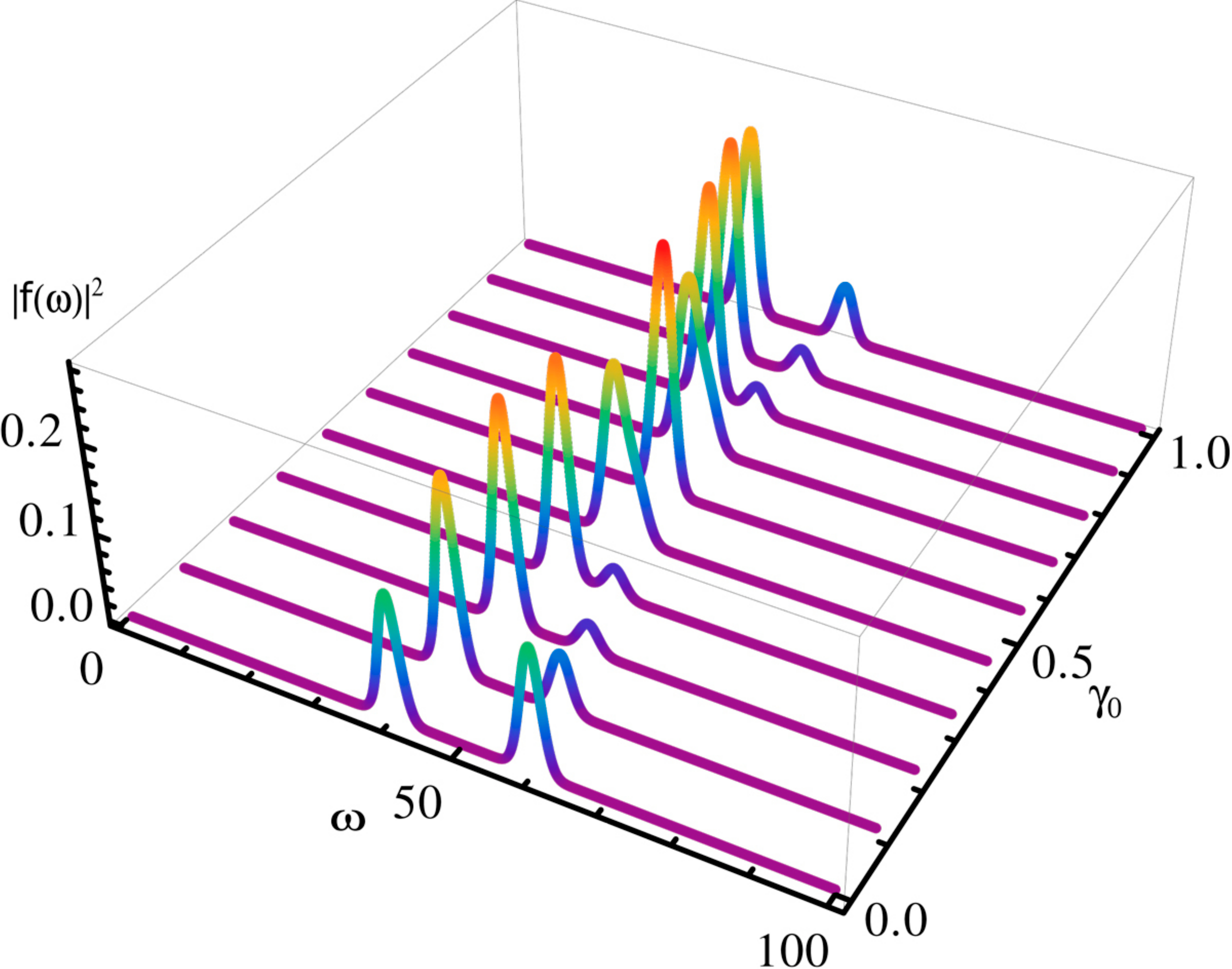}
&\hspace{2cm} &
\includegraphics[width=6cm]{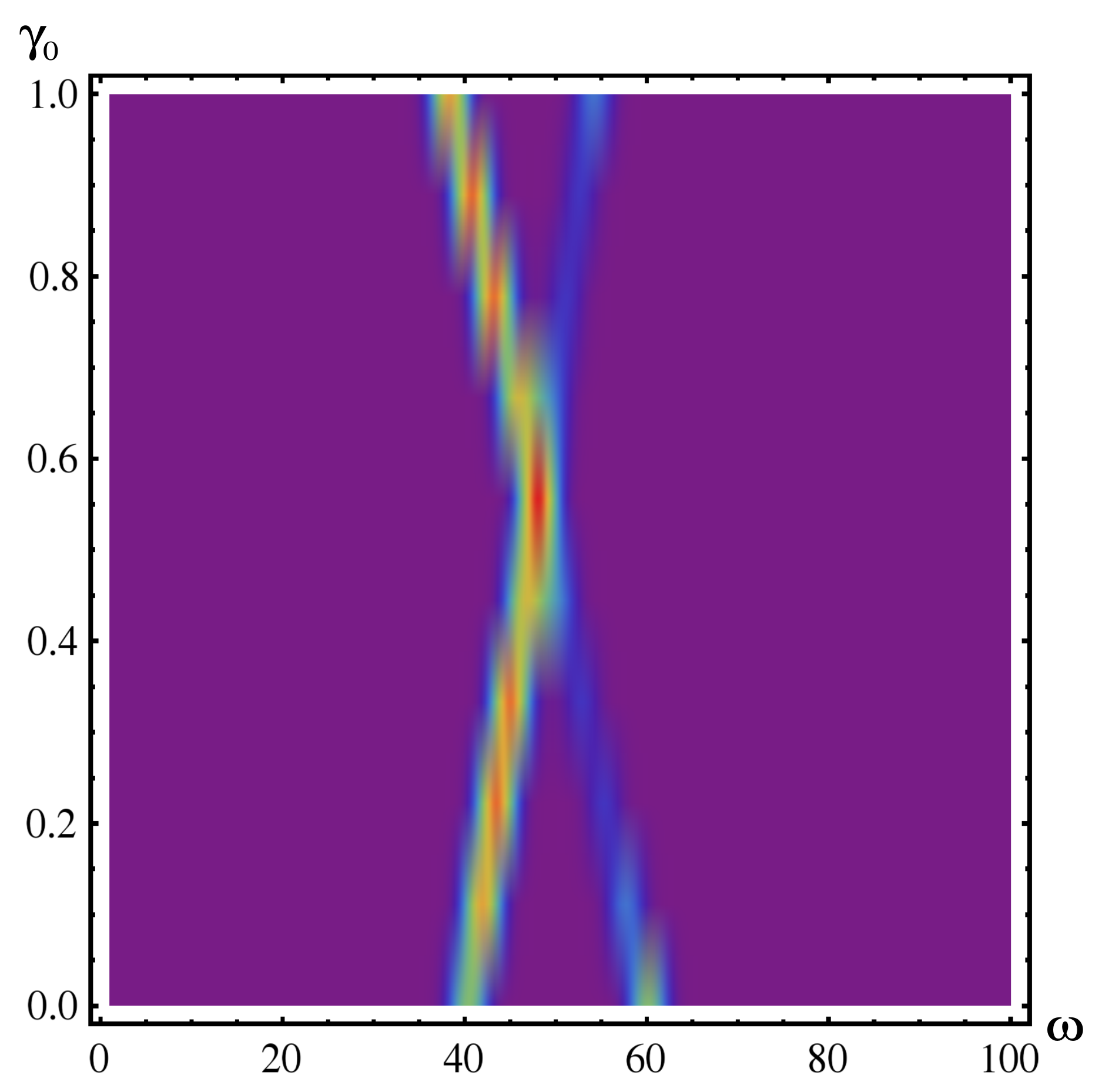}
\end{tabular}
\end{center}
\caption{(Color online). Initial distribution $|f(\omega)|^2$ of photons in the frequency domain as a function of the
control parameter $\gamma_0$. {\bf a)} 3D plot showing the crossing from double picked to single picked initial distributions. {\bf b)} Density plot showing as the height and the positions of the peaks changes as a function of $\gamma_0$.}
\label{fig:fofomega}
\end{figure}

In the following we will calculate the dynamical map and the exact two point correlation functions
for the dynamics of the polarization of a photon (system) interacting with
the spatial components of it's wave packet (environment).

\subsection{Dynamical map}
The dynamical map is easily calculated to be:

\begin{equation}
\begin{aligned}
 \Phi_{0}^{t}[\op{\rho_S}{×}{×}]&=\mathrm{Tr}_E\left[\opt{U}{t}{}{}\op{\rho_S}{×}{×}\otimes\ketbra{\chi}{\chi}\opt{U}{t}{}{\dag}\right]\\
 &=\op{\Pi}{H}{}\op{\rho}{S}{×}\op{\Pi}{H}{}+\op{\Pi}{V}{}\op{\rho}{S}{}\op{\Pi}{V}{}+g^{*}(t)\op{\Pi}{H}{}\op{\rho}{S}{×}\op{\Pi}{V}{}+g(t)\op{\Pi}{V}{}\op{\rho}{S}{×}\op{\Pi}{H}{},\\
 g(t)&=\int d\omega \;|f(\omega)|^{2} e^{-\imath \Delta n \omega t},
\end{aligned}
\end{equation}
where we defined the projectors $\op{\Pi}{H}{}=\ketbra{H}{H}$ and $\op{\Pi}{V}{}=\ketbra{V}{V}$.

It easy to see that the damping basis in this case is given by 
${\op{\Pi}{H}{},\op{\Pi}{V}{},\ketbra{H}{V},\ketbra{V}{H}}$.
The two-point correlation functions calculated by means of the dynamical map only are thus given
by:

\begin{equation}
\begin{aligned}
<\;\opt{\sigma}{t+\tau}{\alpha}{}\opt{\sigma}{t}{\beta}{}\;>_P&=\mathrm{Tr}\bigg[\opt{\sigma}{0}{\alpha}{}\Phi_{0}^{\tau}\opt{\sigma}{0}{\beta}{}\Phi_{0}^{t}\opt{\rho}{0}{S}{}\bigg]\nonumber\\
&= \sum_{k,l,j,i}(B_{\alpha})_{k}^{0}v_{l}^{k}(\tau)(B_{\beta})_{j}^{l} v_{i}^{j}(t)\, c^i(0),
\end{aligned}
\end{equation}
where $(B_{\alpha})_{i}^{j}=\mathrm{Tr}\big[\check{\Lambda}^{j}\op{\sigma}{\alpha}{}\op{\Lambda}{i}{}\big]$

\subsection{Exact two-point correlation functions}

Now we go on with the calculation of the exact two point correlation functions
that we will use to check our definition of non-Markovianity regime.
They are given by:

\begin{equation}
\begin{aligned}
Tr_E\big[\opt{o}{t_1}{1}{}\opt{o}{t_2}{2}{}\,\op{\rho}{E}{}\big]&=\op{\Pi}{H}{}\,\opt{o}{0}{1}{}\,\op{\Pi}{H}{}\,\opt{o}{0}{2}{}\op{\Pi}{H}{}+\op{\Pi}{V}{}\,\opt{o}{0}{1}{}\,\op{\Pi}{V}{}\,\opt{o}{0}{2}{}\op{\Pi}{V}{}\\
&+g(t_1)\op{\Pi}{H}{}\,\opt{o}{0}{1}{}\,\op{\Pi}{V}{}\,\opt{o}{0}{2}{}\op{\Pi}{V}{}+g(-t_1)\op{\Pi}{V}{}\,\opt{o}{0}{1}{}\,\op{\Pi}{H}{}\,\opt{o}{0}{2}{}\op{\Pi}{H}{}\\
&+g(t_2)\op{\Pi}{H}{}\,\opt{o}{0}{1}{}\,\op{\Pi}{H}{}\,\opt{o}{0}{2}{}\op{\Pi}{V}{}+g(-t_2)\op{\Pi}{V}{}\,\opt{o}{0}{1}{}\,\op{\Pi}{V}{}\,\opt{o}{0}{2}{}\op{\Pi}{H}{}\\
&+g(t_1-t_2)\op{\Pi}{H}{}\,\opt{o}{0}{1}{}\,\op{\Pi}{V}{}\,\opt{o}{0}{2}{}\op{\Pi}{H}{}+g(t_2-t_1)\op{\Pi}{V}{}\,\opt{o}{0}{1}{}\,\op{\Pi}{H}{}\,\opt{o}{0}{2}{}\op{\Pi}{V}{}.\\
\end{aligned}
\end{equation}

\newpage
\section{Trace distance measure}
Here we give a brief overview of the trace distance measure \cite{breuer}, which has been widely 
used in order to calculate the non-Markovianity of quantum maps.
In particular we refer to a recent experiment where a system plus environment
dynamics has been simulated using an optical set-up and where it has been shown
that, according to the trace distance measure, it is possible to switch from
Markovian to non-Markovian dynamics by simply changing the distribution in frequency
of the photon's wavepacket.

\subsection{Trace distance based measure}
The trace distance is a metric in the space of (density) matrices, which tells
us how different they are.
The quantification of the non-Markovian behavior through the trace distance relies on
the idea that Markovian dynamics would take {\it any} initial state into an asymptotic
state {\it monotonically}. Here monotonically means that the trace distance between the
initial state and the asymptotic state of the map is a non-increasing function of time.

Since the asymptotic state could be difficult to be found
the measure on non-Markovianity introduced uses the evolution of
two initial states and compares their relative distance along the evolution
under the dynamical map whose behavior has to be checked.

Let us call $\Phi_{0}^{t}$ the dynamical map, $\op{\rho}{1}{}$ and $\op{\rho}{2}{}$ two initial states,
$\mathcal{D}(\op{\rho}{×}{×},\op{\sigma}{×}{×})=\frac{1}{2}\sum\limits_{i}|\chi_i|$ the trace distance
where $\chi_i$ are the eigenvalues of the matrix $\op{\rho}{×}{×}-\op{\sigma}{×}{×}$.

In order to calculate the trace distance one has to follow the following receipt:
\begin{itemize}
\item Calculate the evolved density matrices $\opt{\rho}{t}{1}{×}=\Phi_{0}^{t}[\op{\rho}{1}{}]$
and $\opt{\rho}{t}{2}{×}=\Phi_{0}^{t}[\op{\rho}{2}{}]$;
\item Calculate the trace distance $\mathcal{D}(\opt{\rho}{t}{1}{×},\opt{\rho}{t}{2}{×})$:
\item Calculate the derivative $f(t)=\frac{d}{dt}\mathcal{D}(\opt{\rho}{t}{1}{×},\opt{\rho}{t}{2}{×})$;
\item Calculate the integral $N(\Phi)=\int_{f(t)>0} d\tau f(\tau)$ over intervals where the derivative is positive,
thus signaling an increase in the distance between the two states at that point in time;
\item Maximize over all possible initial states pairs $\op{\rho}{1}{}$ and $\op{\rho}{2}{}$.
\end{itemize}

\begin{figure}[t]
\label{fig_pm}
\begin{center}
\begin{tabular}{ccc}
{\bf a)}&{\bf b)}&{\bf c)}\\
&&\\
\includegraphics[width=5cm]{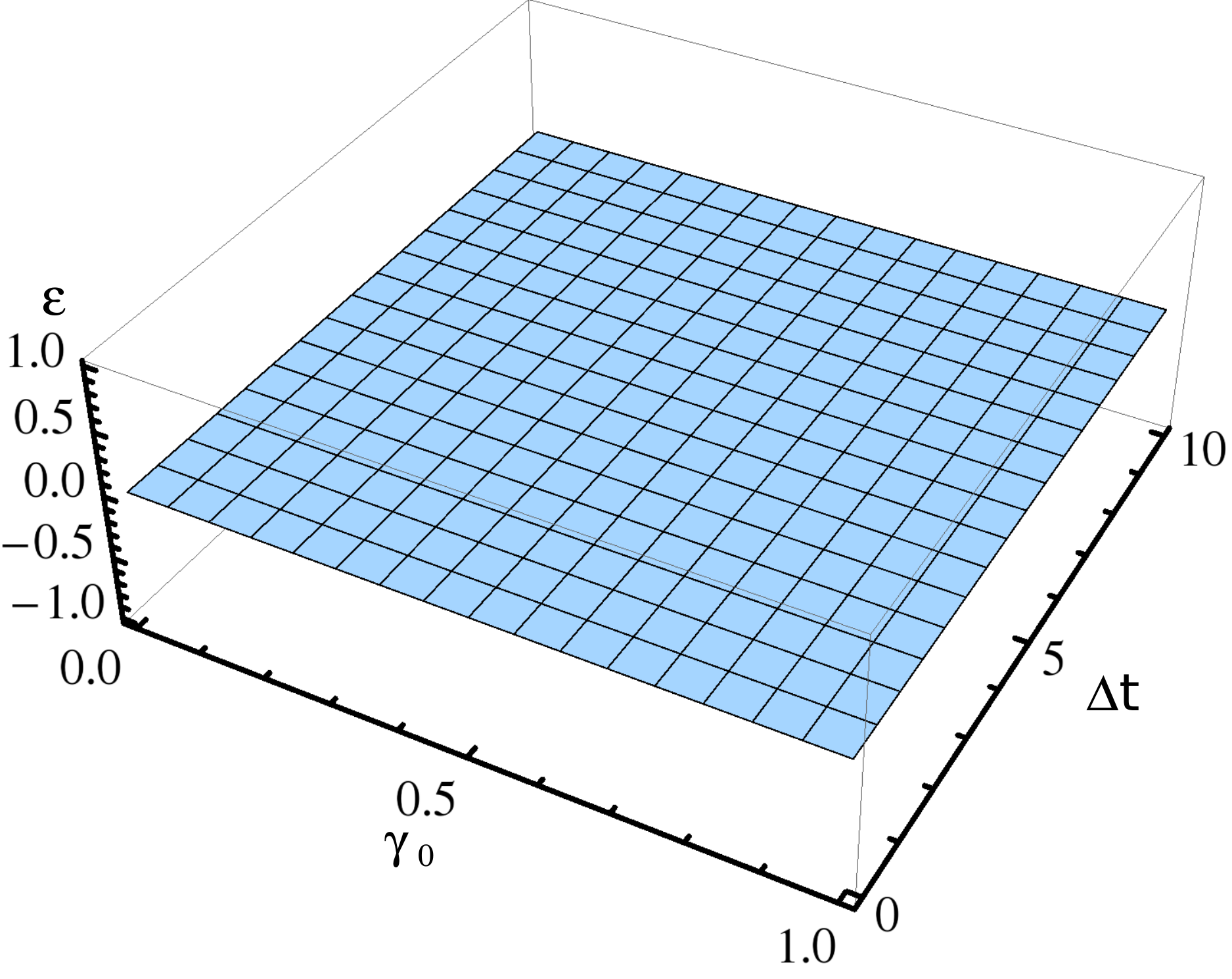}
&
\includegraphics[width=5cm]{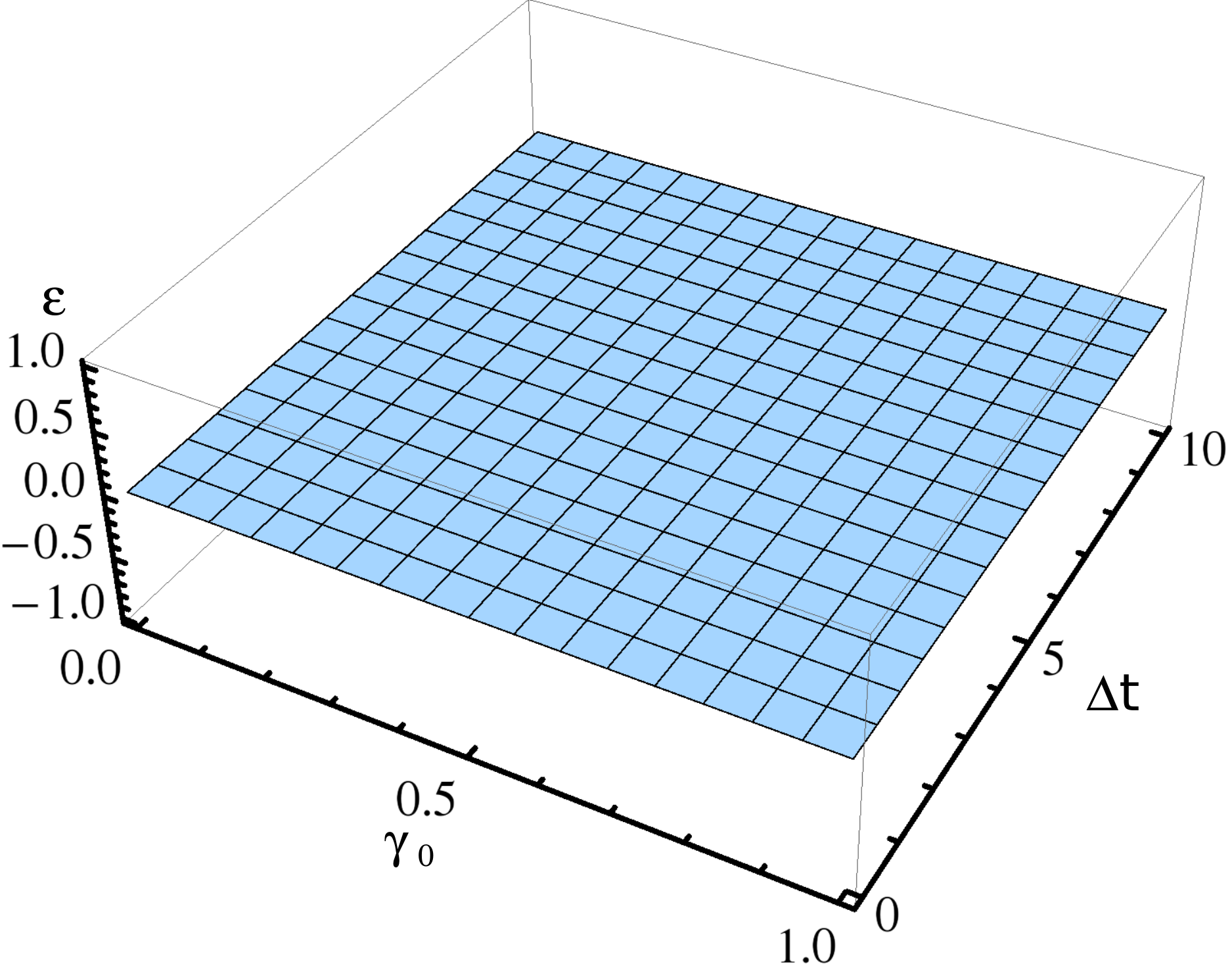}
&
\includegraphics[width=5cm]{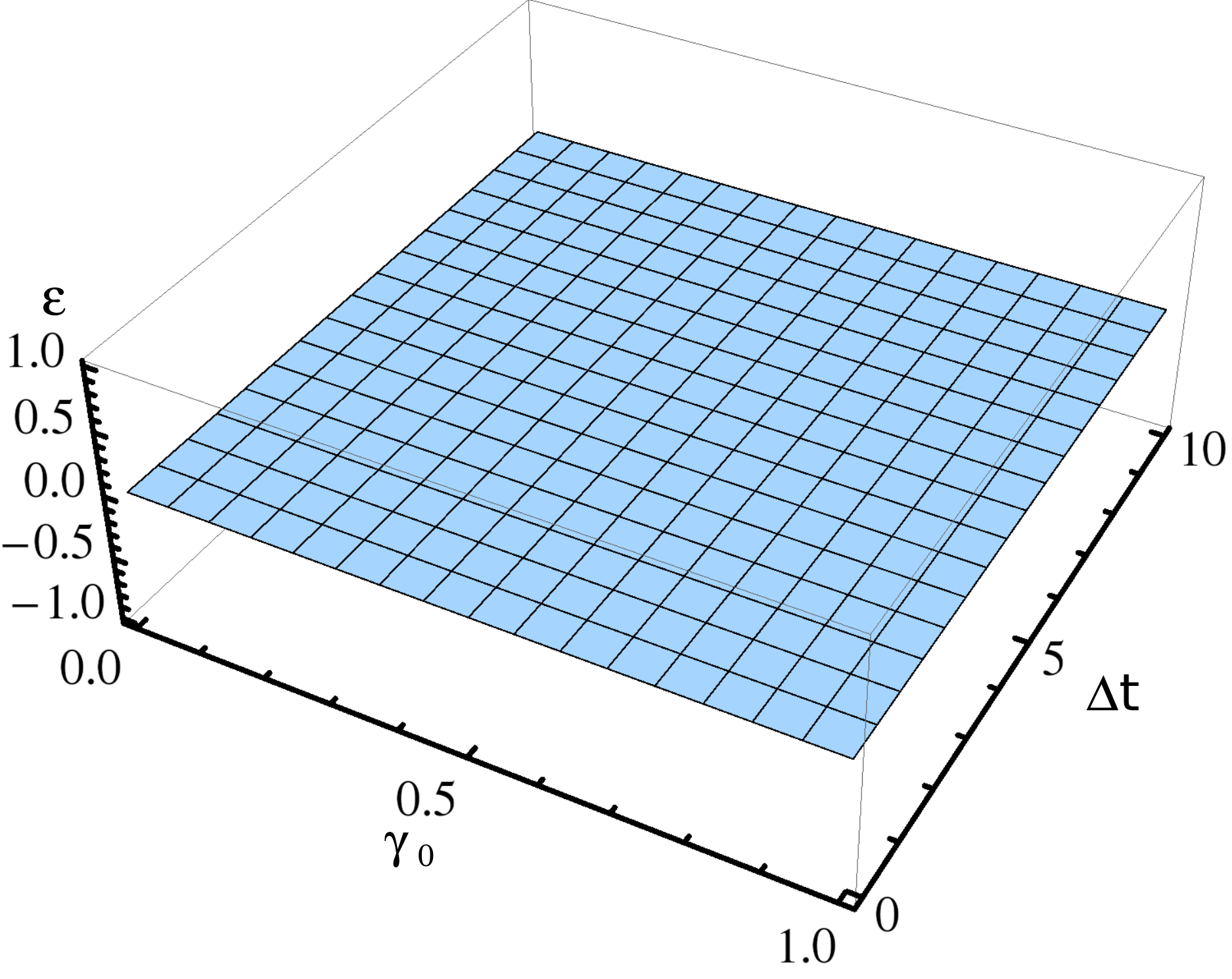}
\end{tabular}
\end{center}
\caption{(Color online). Relative change $\varepsilon$ of the correlation function $<\opt{\sigma}{t+\delta t}{}{z}\opt{\sigma}{t}{}{z}>$ 
as the parameter $\gamma_0$ is changed for the case of {\bf a)} $t=0$, {\bf b)} $t=5$, {\bf c)} $t=10$. Note that the relative change $\varepsilon$ is always zero, which implies that according to the proposed non-Markovianity criteria the process is always Markovian.}
\end{figure}

\begin{figure}[t]
\label{fig_zz}
\begin{center}
\begin{tabular}{ccc}
{\bf a)}&{\bf b)}&{\bf c)}\\
&&\\
\includegraphics[width=5cm]{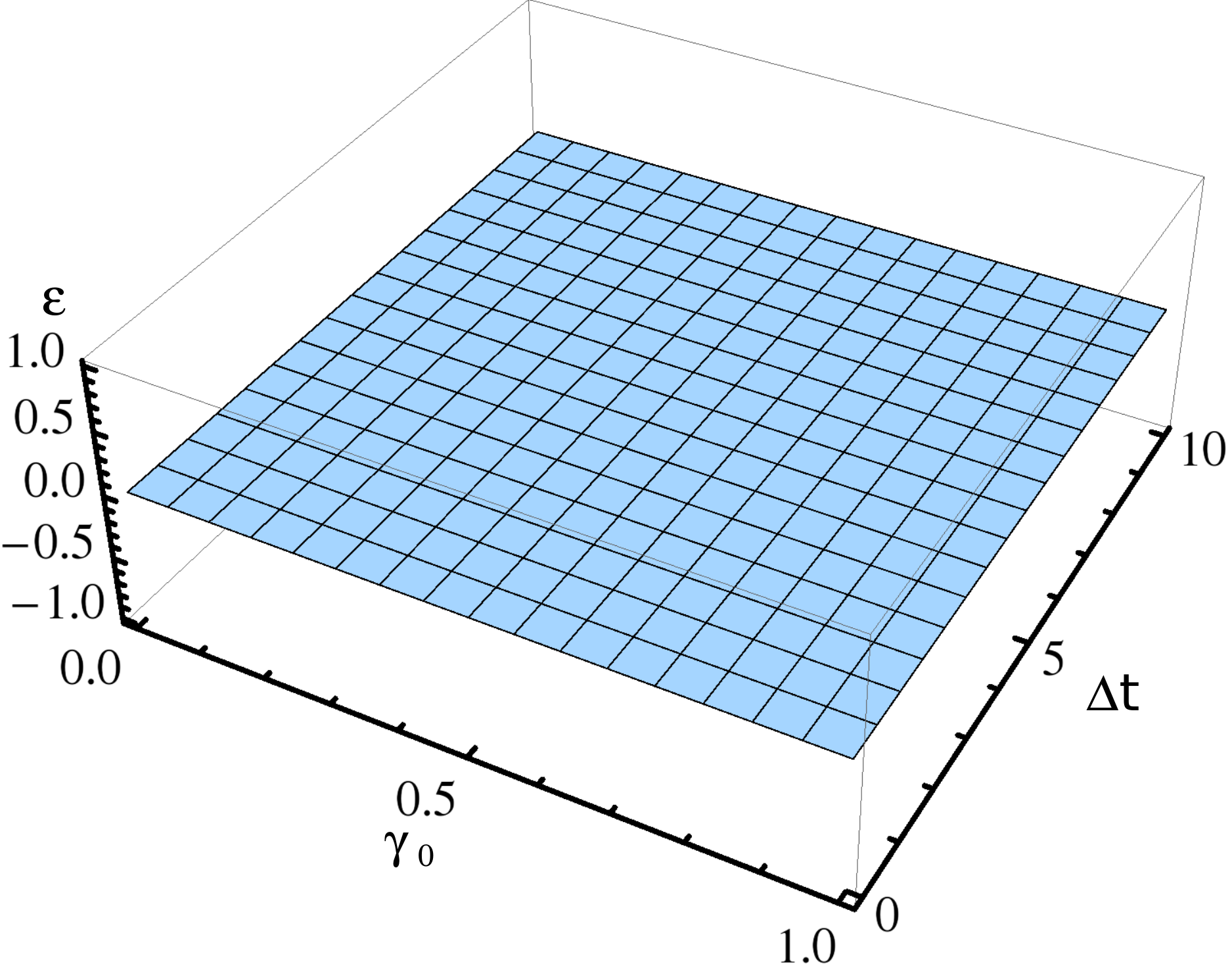}
&
\includegraphics[width=5cm]{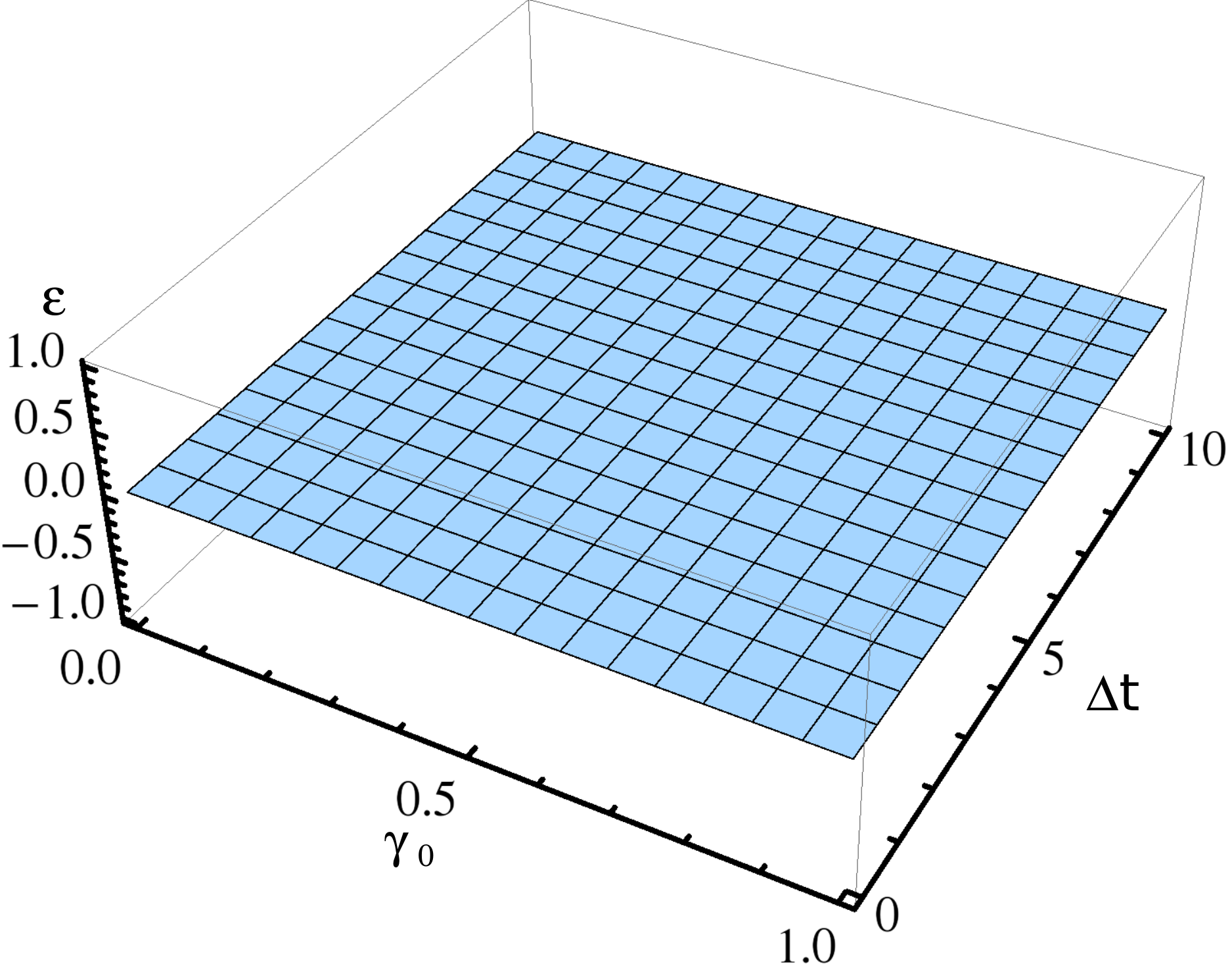}
&
\includegraphics[width=5cm]{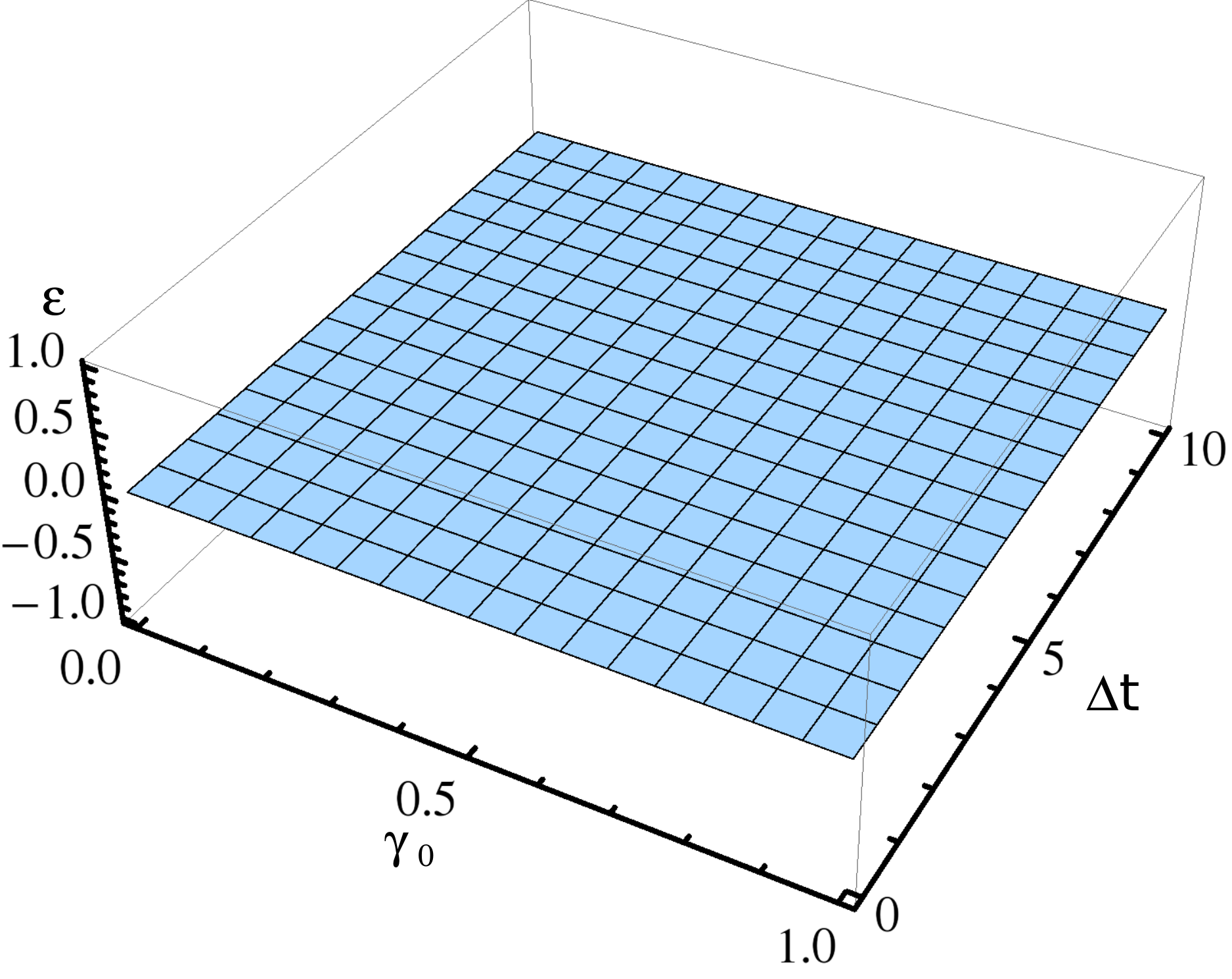}
\end{tabular}
\end{center}
\caption{(Color online). Relative change  $\varepsilon$ of the correlation function $<\opt{\sigma}{t+\delta t}{}{+}\opt{\sigma}{t}{}{-}>$ 
as the parameter $\gamma_0$ is changed for the case of {\bf a)} $t=0$, {\bf b)} $t=5$, {\bf c)} $t=10$.  Note that the relative change $\varepsilon$ is always zero, which implies that according to the proposed non-Markovianity criteria the process is always Markovian.}
\end{figure}

Now we switch to the analysis of the above system according to our definition (see main text)
of Markovian regime.
We looked at the two-point correlation functions $<\opt{\sigma}{t+\Delta t}{}{+}\opt{\sigma}{t}{}{-}>$ and
$<\opt{\sigma}{t+\Delta t}{}{z}\opt{\sigma}{t}{}{z}>$ as a function of both the control parameter $\gamma_0$ 
and $\Delta t$ for different times $t$. In particular we looked at the parameter $\epsilon=1-<\opt{\sigma}{t+\Delta t}{}{\alpha}\opt{\sigma}{t}{}{\beta}>_{M}/<\opt{\sigma}{t+\Delta t}{}{\alpha}\opt{\sigma}{t}{}{\beta}>_{EXP}$.
We can see from Figs. 2 and 3 that the system can be considered to be in a Markov
regime regardless of the value of $\gamma_0$ ($\epsilon=0$), {\it i.e.} regardless of the number of
peaks of the initial distribution $|f(\omega)|^{2}$.

\newpage
\section{Divisibility measure}

In this section we will evaluate the non-Markovianity of the system accordingly to
the so-called divisibility measure ~\cite{div1,div2}.
Moreover we show that it has a link with the classical case 
in which it is possible to show that in general (Markov and non-Markov regimes) one can
construct a set of two-point propagators with the semigroup property.

\subsection{Definition of the divisibility measure}

\noindent We have seen above that in all cases considered the master equation reads:

\begin{equation}
 \frac{d\rho_S(t)}{dt}= \mathcal{L}(t)\rho_S(t),
\end{equation}

\noindent where 

\begin{eqnarray}
\mathcal{L}(t)\opt{\rho}{0}{S}{}&=&-\frac{\imath }{2}\left(\omega_0+\frac{S(t)}{2}\right)[\op{\sigma}{z}{},\opt{\rho}{t}{S}{}]+\gamma(t)\left(\op{\sigma}{-}{}\opt{\rho}{t}{S}{}\op{\sigma}{+}{}-\frac{1}{2}\left\{\op{\sigma}{+}{}\op{\sigma}{-}{},\opt{\rho}{t}{S}{}\right\}\right),\nonumber\\
\mathcal{L}(t)\opt{\rho}{0}{S}{}&=&\frac{h(t)e^{-h(t)}}{2}\left(\op{\sigma}{z}{}\,\opt{\rho}{0}{S}{}\,\op{\sigma}{z}{}-\,\opt{\rho}{0}{S}{}\right),\nonumber\\
\mathcal{L}(t)\opt{\rho}{0}{S}{}&=&\frac{dg^{*}(t)}{dt}\op{\Pi}{H}{}\,\opt{\rho}{0}{S}{}\,\op{\Pi}{V}{}\frac{dg(t)}{dt}\op{\Pi}{V}{}\,\opt{\rho}{0}{S}{}\,\op{\Pi}{H}{},\nonumber
\end{eqnarray}

\noindent for decay, decoherence in a thermal phononic bath and decoherence 
in an engineered environment.

\noindent The dynamical map is thus easily found to be

\begin{equation}
 \Phi_{t_0}^{t}= e^{\int_{t_0}^{t}d\tau\mathcal{L}(\tau)}.
\end{equation}

\noindent The proposed measure is given by 

\begin{eqnarray}
 \mathcal{I}(\Phi)&=&\int_0^{\infty} d\tau\; d(\tau),\nonumber\\
  d(t)&=& {lim}_{\epsilon \rightarrow 0^+} \frac{||(\Phi_{t}^{t+\epsilon}\otimes 1)\ketbra{\Psi}{\Psi}||_{1}-1}{\epsilon},\nonumber\\
  \ket{\Psi}&=&\frac{1}{2}\left(\ket{00}+\ket{11}\right).\nonumber
\end{eqnarray}

\noindent If we now choose a basis for the two qubits which is given by the tensor product of the damping basis 
for the first qubit and the standard basis for the second one we can write

\begin{eqnarray}
  (\Phi_{t}^{t+\epsilon}\otimes 1)\ketbra{\Psi}{\Psi}&=&\sum_{i,j}(\Phi_{t}^{t+\epsilon}\otimes 1)c_{i,j} \op{\Lambda}{i}{}\otimes \op{\sigma}{j}{},\nonumber\\
  =\sum_{i,j}e^{\int_{t}^{t+\epsilon}d\tau\lambda_i(\tau)}c_{i,j} \op{\Lambda}{i}{}\otimes \op{\sigma}{j}{}&\approx&
  \sum_{i,j}e^{\lambda_i(t)\epsilon}c_{i,j} \op{\Lambda}{i}{}\otimes \op{\sigma}{j}{},\nonumber\\
  c_{i,j}&=& \mathrm{Tr}[\opd{\Lambda}{i}{}\otimes \opd{\sigma}{j}{}\ketbra{\Phi}{\Phi}].
\end{eqnarray}

\subsection{Relation to the classical case}

\noindent We now turn to the question: is it always possible to construct a divisible map which gives the right
mean values as in the classical case?
Let us try to follow the argument by H\"anggi and Thomas \cite{hanggi1,hanggi2}.
First we assume that we have solved the problem of finding the dynamical map $\mathcal{E}_{(t,t_0)}$, 
which evolves the initial state  $\opt{\rho}{t_0}{S}{}$ of a system from $t_0$ to $t$.
In general we cannot write $(t_2\ge t_1\ge t_0)$:

\begin{equation}
 \rho_S(t_2+t_1)=\Phi_{t_1}^{t_2}\Phi_{t_0}^{t_1}\opt{\rho}{t_0}{S}{}=\Phi_{t_1}^{t_2}\opt{\rho}{t_1}{S}{}
\end{equation}

\noindent simply because we do not know how to evolve a state starting from $t_1$ up to $t_2$. Is it possible to find a dynamical map which allows us to do so?

We first define the superoperator

\begin{equation}
 \mathcal{G}(t)=\frac{d\Phi_{t_0}^{s}}{ds}\bigg|_{s=t^+} {\Phi^{-1}}_{t_0}^{t}.
\end{equation}

\noindent The inverse evolution ${\Phi^{-1}}_{t_0}^{t}$ acts in such a way that 
the generator is independent of the initial state, exactly as in the classical case.
The above superoperator is well defined since $\mathcal{E}_{(s,t_0)}$ is
known by assumption and ${\Phi^{-1}}_{t_0}^{t}$ is meaningful since we know that 
the ``forward evolution" $\Phi_{t_0}^{t_1}$ started from a meaningful state.
Thus $\mathcal{G}(t)$ represents the infinitesimal generator of a
dynamical map, which evolves the system from $t$ to $t^+$.
The dynamical map we are looking for thus satisfies the equation:

\begin{equation}
 \frac{d\mathcal{D}_{t_1}^{t}}{dt}=\mathcal{G}(t)\mathcal{D}_{t_1}^{t},
\end{equation}

\noindent the choice of $t_1$ as the second argument is due to the fact that
we want $\opt{\rho}{t_1}{S}{}$ to be the initial state of the interrupted evolution.
Exactly as in the classical case by changing the initial time $t_1\ge t_0$
it is possible to find a whole set of dynamical maps:

\begin{equation}
 \mathcal{D}_{t_1}^{t}=\text{T}e^{\int_{t_1}^{t}d\tau \;\mathcal{G}(\tau)}.
\end{equation}

The form of the new dynamical maps makes it clear that all of them belong
to a semigroup, namely ($t_2\ge t_1$):
\begin{equation}
 \mathcal{D}_{t_1}^{t_2}=\mathcal{D}_{s}^{t_2}\mathcal{D}_{t_1}^{s}.
\end{equation}

We stress that as in the classical case the new dynamical map will
give us only the right mean values, but not the correlation functions.
So in fact we built a dynamical map for a Markov process which has the same mean 
values of the original (possibly) non-Markovian process but with different correlation functions.

By using the damping basis defined above it is easy to define the inverse dynamical map:

\begin{eqnarray}
  \opt{\rho}{t_0}{S}{}={\Phi^{-1}}_{t_0}^{t}\opt{\rho}{t}{S}{}&=&\sum_{i}{\Phi^{-1}}_{t_0}^{t}c_{i}(t) \op{\Lambda}{i}{},\nonumber\\
  =\sum_{i}e^{\int_{t_0}^{t}d\tau\lambda_i(\tau)}c_{i}(t_0) {\Phi^{-1}}_{t_0}^{t}\op{\Lambda}{i}{}&=&\sum_{i}e^{\int_{t_0}^{t}d\tau\lambda_i(\tau)}c_{i}(t_0) e^{-\int_{t_0}^{t}d\tau\lambda_i(\tau)}\op{\Lambda}{i}{},\nonumber\\
  &=&\sum_{i}c_{i}(t_0)\op{\Lambda}{i}{},\nonumber
\end{eqnarray}

\noindent where we defined the inverse map by its action on the basis elements ${\Phi^{-1}}_{t_0}^{t}\op{\Lambda}{i}{}= e^{-\int_{t_0}^{t}d\tau\lambda_i(\tau)}\op{\Lambda}{i}{}$.

In this particular case where the damping basis is time independent we can thus write

\begin{equation}
\label{eq:infevol}
\Phi_{t}^{t+\epsilon}=\Phi_{t_0}^{t+\epsilon}{\Phi^{-1}}_{t_0}^{t}=e^{\int_{t_0}^{t+\epsilon}d\tau \;\mathcal{L}(\tau)}e^{-\int_{t_0}^{t}d\tau \;\mathcal{L}(\tau)}=e^{\int_{t}^{t+\epsilon}d\tau \;\mathcal{L}(\tau)}.
\end{equation}

The above equation is very important for our purposes because it shows that the 
infinitesimal generator of the actual dynamics, namely $\mathcal{L}(t)$,
is also the generator of the {\it ad-hoc} constructed divisible map $\mathcal{G}(t)\equiv \mathcal{L}(t)$.
The dynamical maps $\Phi$ has already the semigroup property.

But in Eq. (\ref{eq:infevol}) we can also recognize the ground of the divisibility
measure. It tests whether the infinitesimal evolution from any point $t$ up to $t+\epsilon$
is a meaningful one.
Nevertheless, we have just shown that it is possible to trick this measure by constructing an
{\it ad-hoc} map $\mathcal{D}$ by the knowledge of the map $\Phi_{t_0}^{t}$ and the initial
state $\opt{\rho}{t_0}{S}{}$.
\end{widetext}


\begin{thebibliography}{99}
\bibitem{calc}It is easy to show that any term of the form
$\text{Tr}_{E}\left[\opt{U}{t_2,t_1}{}{}\;\op{o}{}{}\otimes \op{\rho}{E}{}\;\opt{U}{t_2,t_1}{}{\dag}\right]$ 
can be rewritten as $\Phi_{t_2,t_1}[\op{o}{}{}]$. This is done by using trace-less operators 
plus the identity as basis for both system's and environment's Hilbert spaces.
\end{thebibliography}
\end{document}